\documentclass[11pt,amsmath,amssymb,nofootinbib]{revtex4-2} 
\usepackage{graphicx}
\usepackage{dcolumn}
\usepackage{bm}

\usepackage[utf8]{inputenc}
\usepackage[T1]{fontenc}
\usepackage{mathptmx}
\usepackage{etoolbox}

\makeatletter
\def\@email#1#2{%
 \endgroup
 \patchcmd{\titleblock@produce}
  {\frontmatter@RRAPformat}
  {\frontmatter@RRAPformat{\produce@RRAP{*#1\href{mailto:#2}{#2}}}\frontmatter@RRAPformat}
  {}{}
}%
\makeatother

\usepackage{xcolor}

\usepackage{soul} 

\newcommand{\commentout}[1]{}

\newcommand{\RevOut}[1]{{}}

\newcommand{\OhNum}{{\mathrm{Oh}}}
\newcommand{\kB}{{k_\mathrm{B}}}
\newcommand{\InitRad}{R_0}

\newcommand{\SpeciesFlux}{\mathcal{F}}
\newcommand{\ReversibleStress}{\mathcal{R}}
\newcommand{\WhiteNoiseMass}{\mathcal{Z}}
\newcommand{\WhiteNoiseMomentum}{\mathcal{W}}
\newcommand{\ViscousTensor}{\boldsymbol{\tau}}

\newcommand{\Isp}{{\mathrm{I}}}    
\newcommand{\Osp}{{\mathrm{O}}}
\newcommand{\OAsp}{{\mathrm{O}*}}
\newcommand{\Ssp}{{\mathrm{S}}}

\begin{document}

\preprint{AIP/123-QED}

\title[Fluctuating Hydrodynamics and Surfactant-laden Interfaces]{A Fluctuating Hydrodynamics Model for Nanoscale Surfactant-laden Interfaces}

\author{John B. Bell*}\email{jbbell@lbl.gov} 
\author{Andrew Nonaka}%
\affiliation{ 
Lawrence Berkeley National Laboratory, Berkeley, California 94720 USA
}%

\author{Alejandro L. Garcia}
\affiliation{%
Dept. of Physics \& Astronomy, San Jose State Univ., San Jose, California 95192 USA}%

\date{\today}

\begin{abstract}
A multispecies diffuse interface model is formulated in a fluctuating hydrodynamics framework for the purpose of simulating surfactant interfaces at the nanoscale.
The model generalizes previous work to ternary mixtures, employing a Cahn-Hilliard free energy density combined with incompressible, isothermal fluctuating hydrodynamics where dissipative fluxes include both deterministic and stochastic terms. 
The intermolecular parameters in the free energy are chosen such that one species acts as a partially miscible surfactant.
From Laplace pressure measurements we show that in this model the surface tension decreases linearly with surfactant concentration, leading to Marangoni convection for interfaces with concentration gradients. 
In the capillary wave spectrum for interfaces with and without surfactant we find that for the former the spectrum deviates significantly from classical capillary wave theory, presumably due to Gibbs elasticity. 
In non-equilibrium simulations of the Rayleigh-Plateau instability, deterministic simulations showed that  the surfactant delays pinching of a fluid cylinder into droplets.
However, stochastic simulations indicate that thermal fluctuations disrupt the surfactant's stabilizing effect. 
Similarly, the spreading of a patch of surfactant, driven by Marangoni convection, was found to be partially suppressed by thermal fluctuations.
\end{abstract}

\maketitle

\section{Introduction}

Surfactants are surface-active compounds, such as oils and soaps, that possess hydrophilic and hydrophobic moieties. 
This unique molecular architecture enables their preferential adsorption at fluid interfaces leading to a significant reduction in surface tension. 
This reduction in interfacial energy is not just a descriptive property but a fundamental thermodynamic principle that reduces the energetic cost associated with creating or maintaining fluid interfaces. 
The dynamic interplay between these surface-active molecules and fluid interfaces defines the complex field of "surfactant-laden flows," which investigates the intricate coupling between fluid mechanics and the behavior of these interfacial agents.\cite{brenner2013interfacial}
Accurately characterizing these systems requires a multi-physics approach for modeling, coupling the Navier-Stokes equations for fluid motion with transport equations that describe surfactant advection, diffusion, and stresses at the interface. 
The study of surfactant-laden flows is of interest across a broad spectrum of scientific and engineering disciplines, impacting critical processes in industrial technology (e.g., coatings, drying, inkjet printing, oil recovery, culinary science) , biological systems (e.g., pulmonary surfactants, tear film dynamics) , and environmental applications (e.g., oil spill dispersion and remediation).
\cite{levich1969surface,craster2009dynamics, bonn2009wetting,lohse2022inkjetreview,mathijssen2023culinary}

Microscopically, fluids are a discrete physical system consisting of molecules that are in constant random motion such that an accurate nanoscale continuum description requires the use of fluctuating fields. 
These thermal fluctuations are accurately described by statistical mechanics \cite{landau_statistical_1980} and they result in interesting and important phenomena, such as Brownian motion and Rayleigh light scattering.
In order to accurately introduce thermal fluctuations in continuum fluid dynamics, we use \emph{fluctuating hydrodynamics} (FHD), originally proposed in linearized form by Landau and Lifshitz.\cite{Landau1959Fluid,LLNS_FD_Fox,Boltzmann_FD_Fox,PhysRev.187.267,ZarateBook2006}. 
The nonlinear hydrodynamic fluctuations were later justified by deriving the Fokker-Planck equations of the distribution function of conserved hydrodynamic quantities \cite{zubarev1983statistical}, which then led to the formulation of the associated stochastic differential equations  (SPDEs) \cite{LLNS_Espanol}.

A variety of diffuse-interface models have been proposed for the study of surfactant-laden flows.\cite{yamashita2024conservative,engblom2013diffuse,zhu2019numerical}
Here we adopt a multispecies formulation based on the Cahn-Hillard model\cite{CahnHilliard:1958,anderson1998diffuse} for phase separated, binary liquid mixtures.
The intermolecular interaction parameters are chosen such that one species plays the role of a partially miscible surfactant. 
The resulting free energy provides the thermodynamic properties of the fluid mixture while the fluid dynamic transport is given by incompressible fluctuating hydrodynamics.

Section~\ref{sec:FHD} presents the details of our FHD model as well as a brief synopsis of the corresponding numerical scheme. 
The method is demonstrated in a variety of examples for equilibrium and nonequilibrium systems in Sections~\ref{sec:EquilibriumSystems} and \ref{sec:NonequilibriumSystems}.  
In these examples we highlight the impact of thermal fluctuations on surfactant-driven flows, summarizing our findings and suggesting avenues for future study in Section~\ref{sec:Conclusions}.



\section{Fluctuating Hydrodynamics}\label{sec:FHD}

This section describes the theoretical and numerical formulations of our fluctuating hydrodynamics model for surfactant-laden flows.
This formulation of multispecies fluctuating hydrodynamics is similar to that described in Barker \textit{et al.}\cite{BrynRP2023} and Bell \textit{et al.}\cite{Bell2025Droplets} extended beyond binary mixtures.
In this paper we consider ternary mixtures with one of the species playing the role of a surfactant.

\subsection{Theory}

We call $c_\alpha$ the concentration for species $\alpha $; for simplicity, we take the molecular mass, $m$, as the same for all species.
Consider a fluid with a free energy density, $\mathcal{G}$, given by\cite{CahnHilliard:1958}
\begin{equation}
\mathcal{G}
= n k_B T \left\{
\sum_{\alpha} c_\alpha \ln c_\alpha 
+ \frac{1}{2} \sum_{\alpha,\beta} \chi_{\alpha\beta} c_\alpha c_\beta 
- \frac{1}{2} \sum_{\alpha,\beta} \kappa_{\alpha\beta} \nabla c_\alpha \cdot \nabla c_\beta
\right\}
\label{eq:FreeEnergyDensity}
\end{equation}
where $n$ is number density, $T$ is temperature, and $k_B$ is the Boltzmann constant.
The Flory coefficients, $\chi$, are determined from the species' interaction energies. 
For $\chi_{\alpha\beta} < 0$ the two species attract each other and favor mixing; when $\chi_{\alpha\beta} > 0$ the interaction is repulsive. 
When $\chi_{\alpha\beta} > 2$ the enthalpy overcomes the entropy of mixing and phase separation occurs between these two species.

For example, a two-species ($\alpha-\beta$) mixture separates into concentrations $c_{e,\alpha}$ and $c_{e,\beta} = 1 - c_{e,\alpha}$ given by solutions of
\begin{align}
\ln \left( \frac{c_e}{1-c_e}\right) = \chi_{\alpha\beta} (2 c_e - 1) 
\label{eq:ce_eqn}
\end{align}
where $c_e$ is either $c_{e,\alpha}$ or $c_{e,\beta}$.
The interfacial coefficient, $\kappa_{\alpha\beta}$, gives a contribution to the free energy due to the interfacial forces in that the gradients increase the free energy when $\kappa_{\alpha\beta} > 0$.
Note that $\boldsymbol{\chi}$ and $\boldsymbol{\kappa}$ are symmetric matrices with $\chi_{\alpha\alpha} = \kappa_{\alpha\alpha}= 0$.
The expected interface thickness is
\begin{align}
    \ell = \sqrt{2}\ell_\mathrm{c} \left(-1-\frac{2\log{4c_e(1-c_e)}}{\chi(1-2c_e)^2}\right)^{-1/2},
    \label{eq:InterfaceThickness}
\end{align}
where $\ell_\mathrm{c} = \sqrt{2\kappa_{\alpha\beta} / \chi_{\alpha\beta}}$ is the characteristic length scale for an interface between species $\alpha$ and $\beta$.
The surface tension is
\begin{align}
    \gamma_{\alpha\beta} = n \kB T \sqrt{2 \chi_{\alpha\beta} \kappa_{\alpha\beta}} ~\sigma_r(\chi_{\alpha\beta})
    \label{eq:SurfaceTension}
\end{align}
where
\begin{align}
    \sigma_r = \int_{c_{e,\alpha}}^{c_{e,\beta}} dc 
    \Big[  \frac{2c}{\chi_{\alpha\beta}} \ln \frac{c}{c_e} + \frac{2c}{\chi_{\alpha\beta}} \ln \frac{1-c}{c_e} - 2 (c - c_e)^2 \Big]^{1/2}
\end{align}
Note that $\gamma_{\alpha\beta} \rightarrow 0$ as either $\chi_{\alpha\beta}\rightarrow 2$ or $\kappa_{\alpha\beta}\rightarrow 0$. 

For systems with more than two species, closed form expressions for interface thickness and interfacial tension are not available.  
For these cases, we have used deterministic numerical simulations of two-dimensional droplets (see Section~\ref{sec:LaplacePressure}) to compute interfacial tension from the Laplace pressure.

The incompressible flow equations for constant density are
\begin{align}
\partial_t( \rho c_\alpha) + \nabla \cdot(\rho u c_\alpha) =& \nabla \cdot \SpeciesFlux_\alpha   \nonumber \\
( \rho u)_t + \nabla \cdot(\rho u u) + \nabla \pi =& \nabla \cdot {\ViscousTensor}  + \nabla \cdot \ReversibleStress \nonumber \\
\nabla \cdot u =& 0
\label{eq:low_mach_eqs}
\end{align}
where $\rho = m n$ is the mass density, $u$ is the fluid velocity and $\pi$ is a perturbational pressure that enforces the incompressibility constraint.
Here, $\SpeciesFlux$, $\ViscousTensor$, and $\ReversibleStress$ are the species flux, the viscous stress tensor, and the reversible stress due to the interfacial tension, respectively.
Although we have formally written the governing equations as stochastic partial differential equations (SPDEs), the equations are too irregular to have a well-defined meaning as SPDEs\cite{hairer2014theory,gubinelli2015paracontrolled}.  Constructing a well-defined mathematical model requires the introduction of a high-wave number cut-off, which is done here by discretizing the system on a finite-sized mesh.

In fluctuating hydrodynamics the dissipative fluxes are written as the sum of deterministic and stochastic terms.
From non-equilibrium thermodynamics, the deterministic species diffusion is given by an Onsager matrix, $\boldsymbol{L}$, multiplied by the thermodynamic driving force 
\begin{align}
\boldsymbol{X} = \frac{\nabla \mu}{T} = \frac{1}{T} \nabla \left[\frac{\delta \mathcal{G}}{\delta c}\right]
\end{align}
and for computational purposes we write it in Fickian form as
\begin{align}
\label{eq:DetSpeciesFlux}
\overline{\SpeciesFlux} = \boldsymbol{L} \boldsymbol{X} = \rho \mathcal{C} \mathcal{D}
\left ( \nabla c +\mathcal{C} \left[ \nabla (\chi c)
+ \nabla (\nabla \cdot \kappa \nabla c\right] \right )
\end{align}
where $\mathcal{C}$ is a diagonal matrix of concentrations and $\mathcal{D}$ is the matrix of species diffusion coefficients.
The Onsager matrix can be expressed in terms of Fickian diffusion matrix as
\begin{align}
\boldsymbol{L} = \frac{\rho m}{k_B} \mathcal{CDC}
\end{align}
The stochastic species diffusion flux can then be written as
\begin{align}
\label{eq:StochSpeciesFlux}
\widetilde{\SpeciesFlux} = 
\sqrt{2 k_B \boldsymbol{L}} \mathcal{Z} = \sqrt{2 \rho m \mathcal{CDC}} \mathcal{Z} = \sqrt{2 \rho m} \mathcal{C} \mathcal{D}^{\frac{1}{2}} \mathcal{Z}
\end{align}
where $\WhiteNoiseMass(\mathbf{r},t)$ is a collection of uncorrelated Gaussian random vector fields with covariance $\delta_{i,j} \delta(t-t') \delta (r-r')$. 
In practice, the matrix $\mathcal{D}^{1/2}$ can be calculated using a Cholesky factorization.
The fourth-order operator in the species equations requires two boundary conditions at solid walls. One of these boundary conditions zeros the net flux through the wall by setting $\overline{\SpeciesFlux} + \widetilde{\SpeciesFlux} = 0$. 
Here we take walls to have neutral wettability for the different components and also set the normal derivative of concentrations to zero.
When there are wettability effects, the normal derivative of concentration at the wall is a nonlinear function of concentration, which is derived from a surface free energy term \cite{ContactAngleBCs2012,Bell2025Droplets}.

The viscous incompressible stress tensor is
$\ViscousTensor =  \overline{\ViscousTensor} +  \widetilde{\ViscousTensor}$
where the deterministic component is
$\overline{\ViscousTensor} = \eta [\nabla u + (\nabla u)^T]$.
The stochastic contribution to the viscous stress tensor is 
$\widetilde\ViscousTensor = \sqrt{\eta \kB T}({\WhiteNoiseMomentum} + {\WhiteNoiseMomentum}^T),
$ where 
${\WhiteNoiseMomentum}(\mathbf{r},t)$ is a standard Gaussian random tensor field
with uncorrelated components.
At walls, the variance of the stochastic fluxes require modification to ensure fluctuation dissipation balance is maintained.  Specifically, on a no-slip wall the viscous stochastic flux is scaled by $\sqrt{2}$ and at a slip wall the viscous stochastic flux is set to zero.\cite{balboa2012staggered}
Finally, the interfacial reversible stress is
\begin{equation}
    \ReversibleStress = \frac{\rho k_B T}{m} \sum_{\alpha, \beta} \left( \nabla c_{\alpha} \otimes \kappa_{\alpha \beta} \nabla c_{\beta} - \frac12 \left( \nabla c_{\alpha} \cdot \kappa_{\alpha\beta} \nabla c_{\beta} \right ) \mathbb{I} \right)
\end{equation}
Since $\ReversibleStress$ is a non-dissipative flux there is no corresponding stochastic flux.
At solid boundaries, we use the normal derivative boundary condition to evaluate $\mathcal{R}$.

Our formulation shares a number of features with deterministic approaches for systems with three or more phases.  
For example, the free energy density, Eq.~\ref{eq:FreeEnergyDensity}, has the same form as that used by Mao \textit{et al.}\cite{mao2019phase}; however, they do not consider coupling with fluid motion.
Our approach is also similar to the formulations of Kim and Lowengrub \cite{kim2005phase}
although their work assumes a diagonal $\kappa$ matrix, which is more restrictive than our approach. In particular, using a diagonal $\kappa$ matrix implies that the interfacial forces between species $\alpha$ and each of the others must be the same, which precludes variation in interfacial forces with composition needed to model surfactants.
Our approach most closely resembles the N-phase model introduced by Dong \cite{DONG2014}, which sets pair-wise nonlinear interfacial coefficients but uses an idealized free energy that is general enough to describe the characteristics of a surfactant but is not suitable for incorporating fluctuations.

\subsection{Numerics}

The numerical scheme is based on methods introduced in earlier work~\cite{donev2014low, Donev_10, RTIL}; details are discussed in Barker \textit{et al.}\cite{BrynRP2023} and its Supporting Information.
The system of equations (\ref{eq:low_mach_eqs}) is discretized using a structured-grid finite-volume approach with cell-averaged concentrations and face-averaged (staggered) velocities with standard spatial discretizations. 
The algorithm uses an explicit discretization of concentration coupled to a semi-implicit discretization of velocity using a predictor-corrector scheme for second-order temporal accuracy.  
The momentum equation is discretized using a Stokes-type splitting.  
Specifically, the advective terms and the reversible stress are computed explicitly using data at time $t^n$ in the predictor and $t^{n+1} = t^n + \Delta t$ in the corrector.  
These terms then form part of the right-hand side of a Stokes system that treats the viscous tensor and the incompressibility constraint implicitly.
The discretized Stokes system is solved by a generalized minimal residual (GMRES) method with a multigrid preconditioner, see Cai \textit{et al.}\cite{cai:2014}.
The explicit treatment of the concentration equation introduces a stability limitation on the time step of
\begin{align}
\lambda_\mathcal{D} \left( \frac{12}{\Delta x^2} + \frac{72}{\Delta x^4}\max_\alpha \;(\sum_\beta \kappa_{\alpha\beta})\; \right) \Delta t \leq 1
\label{eq:StableDt}
\end{align}
where $\lambda_\mathcal{D}$ is the largest eigenvalue of $\mathcal{D}$ and $\Delta x$ is the mesh spacing.

\subsection{Mixture compositions}\label{sec:MixtureCompositions}

We consider two species and three species mixtures composed from combinations of 4 distinct fluid species labeled I, S, O, and O*.
The Flory and interfacial coefficients for these species are listed in Table~\ref{tab:MixtureParameters}.
For droplets and cylinders the inner fluid is always species I and the outer fluid is either species O or O*; by convention, for flat horizontal interfaces, species I is the lower fluid.

In our simulations, we separately considered the following mixtures:
\begin{itemize}
    \item[I-O:] Two-phase mixture of immiscible species, I and O. 
    \item[I-O*:] Two-phase mixture of immiscible species, I and O*, with a surface tension similar to that of an I-S-O mixture interface. 
    \item[I-S-O:] Three-phase mixture consisting of an I-O mixture with an added surfactant species, S, which is immiscible in species O and barely miscible in species I.
\end{itemize}
Species are said to be immiscible when $\chi_{\alpha\beta} > 2$, barely miscible when $\chi_{\alpha\beta} = 2$, and miscible when $\chi_{\alpha\beta} < 2$.

\begin{table}[h]
    \centering
    \begin{tabular}{c|c|c|c|c|}
        ~~ & I-O & I-O* & I-S & O-S \\ \hline
        $\chi_{\alpha\beta}$ & 4.0 & 3.5 & 2.0 & 2.5 \\ \hline
        $\kappa_{\alpha\beta} \times 10^{14}$ ($\mathrm{cm}^2$) & 3.0 & 2.5312 & 1.0 & 2.0 \\ \hline
    \end{tabular}
    \caption{Flory and interfacial coefficients}
    \label{tab:MixtureParameters}
\end{table}

Unless otherwise specified, the physical parameters used in all simulations are as follows:
mass density, $\rho = 1.0~\mathrm{g/cm}^3$, 
molecular mass, $m = 6.0\times 10^{-23}$~g, 
temperature $T = 100$K. 
With these parameters the fluid has approximately 17 molecules per cubic nanometer.
The viscosity is $\eta = 0.01$~poise 
and the binary diffusion diffusion coefficients $\mathcal{D}^\mathrm{bin}_{\alpha\beta}= D = 1.74 \times 10^{-5}~\mathrm{cm^2/s}$ for each $\alpha,\beta$ pair.
With these 
\[
\mathcal{D} = D
\begin{pmatrix}
\frac{1-c_1}{c_1} & -1 &-1 \\
-1 & \frac{1-c_2}{c_2} & -1 \\
-1 & -1 & \frac{1-c_3}{c_3}  
\end{pmatrix}
\]
The equilibrium concentrations for the binary mixtures are:
for I-O: $c_{e,\alpha} = 0.021248$ and $c_{e,\beta} = 1 - c_{e,\alpha} = 0.978752$;
for I-O*: $c_{e,\alpha} = 0.037874$ and $c_{e,\beta} = 1 - c_{e,\alpha} = 0.962126$.

Our simulations had $N_x \times N_y \times N_z$ grid cells; for 2D systems $N_z = 1$. In general $\Delta x = \Delta y = \Delta z = 1$nm, except for the capillary wave system (Section~\ref{sec:CapillaryWave}) where $\Delta x = \Delta y = 1$nm and $\Delta z = 8$nm.
The time step in the simulations was $\Delta t = 0.4$ picoseconds for 2D simulations and $0.2$ps for 3D simulations (see Eq.~(\ref{eq:StableDt})).

\section{Simulation of Equilibrium Systems}\label{sec:EquilibriumSystems}

This section presents simulation results for fluid mixtures, both with and without surfactant, at thermodynamic equilibrium.
First, we describe the measurement of the surface tension as obtained from the Laplace pressure in droplets.
The second scenario is the measurement of surface height fluctuations for a flat interface. 
The spectrum of these fluctuations is analyzed and compared with capillary wave theory.
In the case of a surfactant-laden interface we find that this spectrum deviates significantly from classical theory.

\subsection{Laplace pressure}\label{sec:LaplacePressure}
 
\begin{figure}[h]
    \centering
    \includegraphics[width=0.30\textwidth]{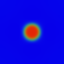}
    \hspace{.2in}
    \includegraphics[width=0.45\textwidth]{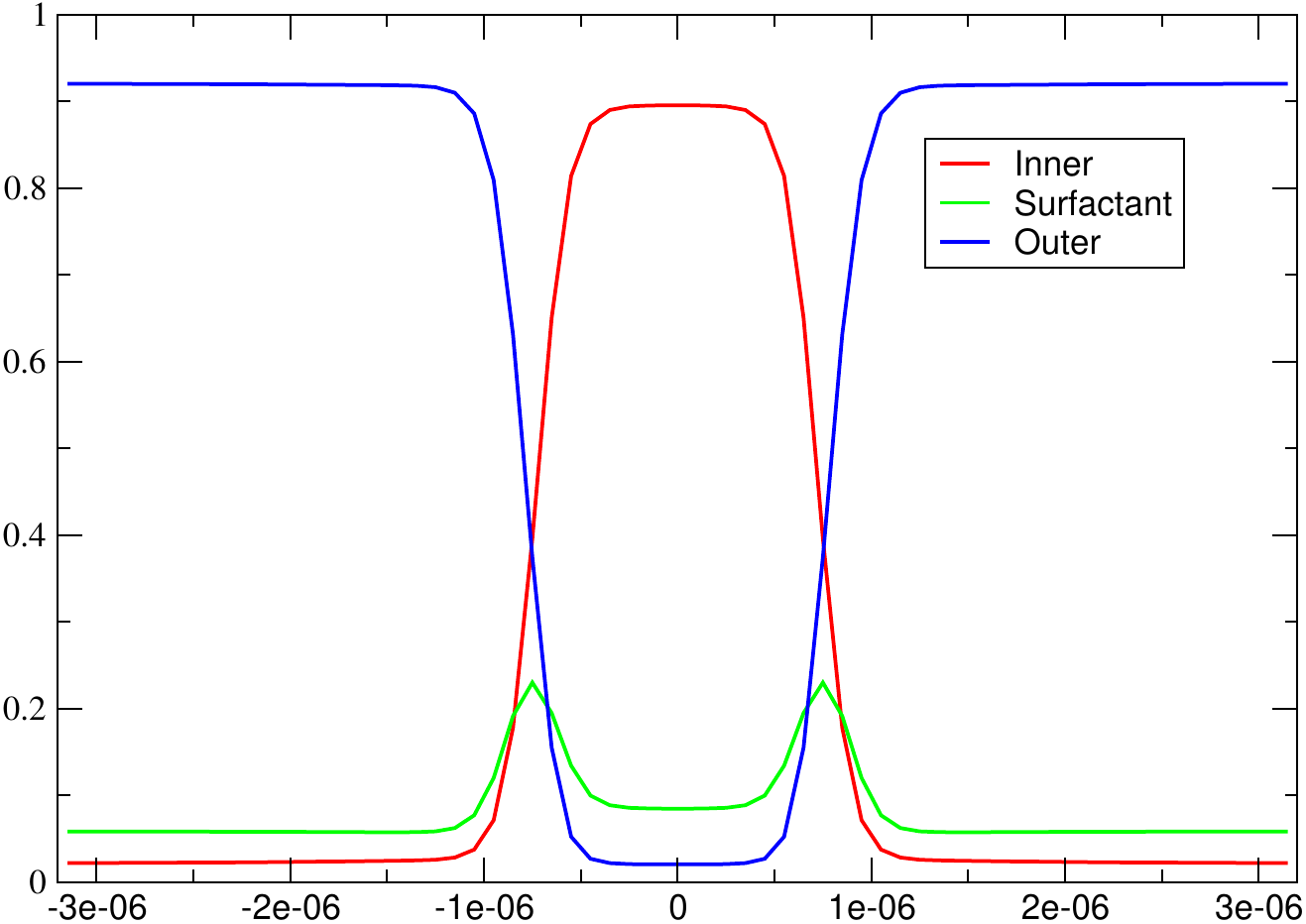}
    \caption{
    (Left) Image of an I-S-O droplet; color map is: Species I (red); Species S (green); Species O (blue).
    (Right) Species concentration profiles at the centerline.
    }
    \label{fig:DropletProfile}
\end{figure}

Deterministic simulations of 2D droplets were run to measure the Laplace pressure $\delta p$.
The concentration profile for an I-S-O droplet is shown in Figure~\ref{fig:DropletProfile} for a simulation domain that is a 64nm square discretized using a $64 \times 64 \times 1$ grid.
Note that the image is created by setting the RGB channel values as $c_\alpha$ 
with: Species I (red); Species S (green); Species O (blue).

Table~\ref{tab:LaplacePressure} shows that for I-O and I-O* mixtures the measured surface tension, $\gamma_{L}= R \delta p$ obtained from the Laplace pressure is in good agreement with the theoretical value given by Eq.~\eqref{eq:SurfaceTension}.
For the case with surfactant (I-S-O), we define the interface (Gibbs dividing surface) as being located where the outer fluid (Species O) reaches concentration $c_\Osp = 0.5$ so the surfactant species is concentrated near the surface yet within the droplet (see Fig.~\ref{fig:DropletProfile}).
For the I-S-O mixture the value of $\gamma_L$ is approximately the same as the I-O* mixture.
This is intentional; the values of $\boldsymbol{\chi}$ and $\boldsymbol{\kappa}$ were specifically chosen such that the two mixtures would have the same surface tension.

\begin{table}[h]
    \centering
    \begin{tabular}{|l|c|c|c|c|}
        \hline
        ~Mixture & $R$ (nm) & $\delta p$ (MPa) & $\gamma_{L}$ (dyne/cm) & $\gamma$ (dyne/cm) \\ \hline
        ~I-O &  8.06 & 40.44  & 32.60 & 31.48 \\ \hline
        ~I-O* & 7.93 & 29.10 & 23.07 & 22.32\\ \hline
        ~I-S-O & 7.99 & 27.90 &  22.29 & -- \\ \hline   
    \end{tabular}
    \caption{Laplace pressure measurements for 2D droplets. 
    In the table, $\gamma_L$ is the measured surface tension and $\gamma$ is the theoretical value given by Eq.~\eqref{eq:SurfaceTension}.}
    \label{tab:LaplacePressure}
\end{table}

\begin{figure}[h]
    \centering
    \includegraphics[width=0.50\textwidth]{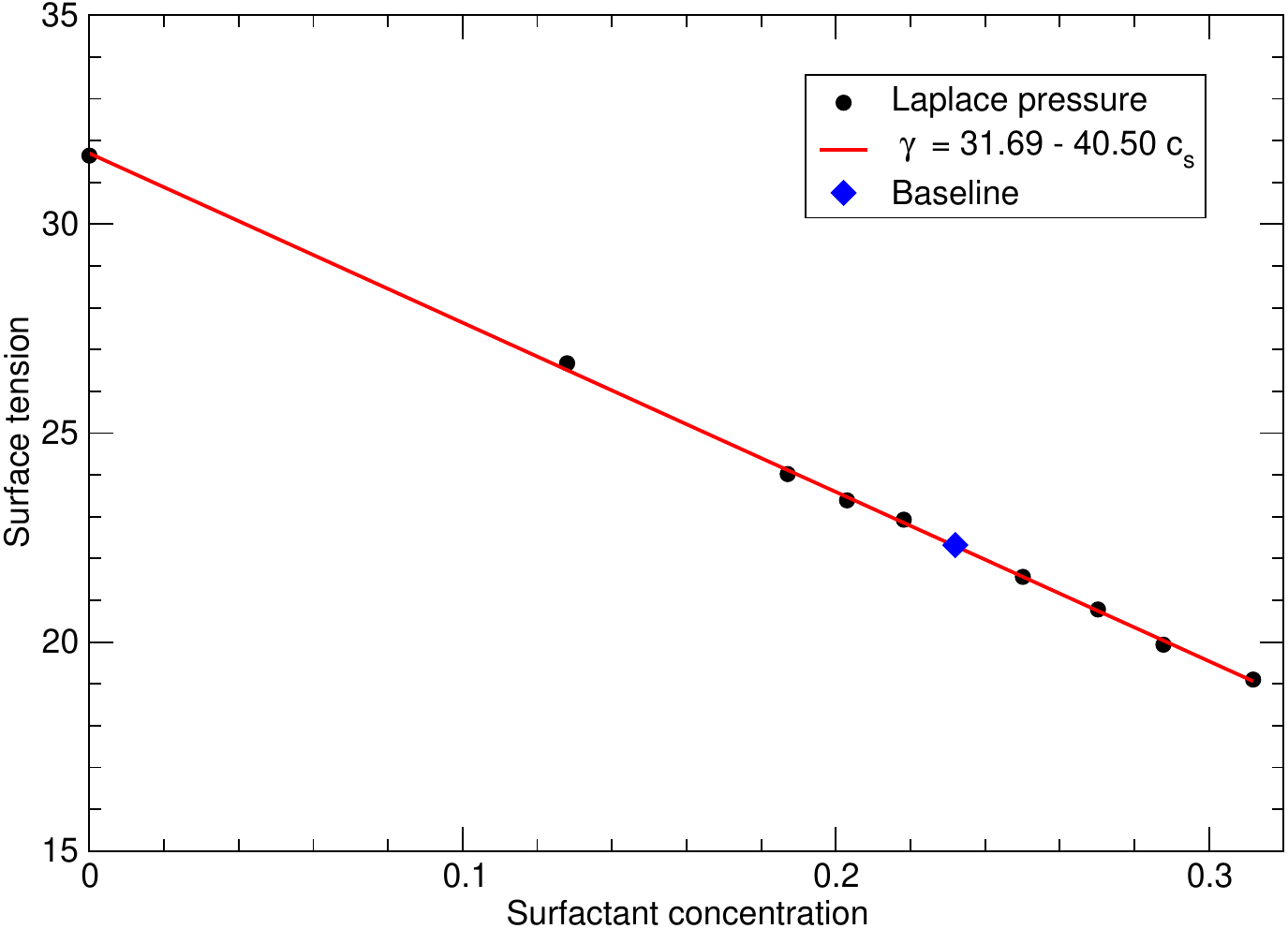}
    \caption{
    Surface tension versus surfactant peak concentration; points are from Laplace pressure measurements and line is the linear fit $\gamma_L(c_\Ssp) = 31.694 - 40.505\, c_\Ssp$ in dynes/cm.
    The baseline value is indicated with a blue diamond.
    }
    \label{fig:GibbsMarangoni}
\end{figure}

The simulations discussed in the sections below used the surfactant concentration of the I-S-O droplet described above; specifically a peak value of $c_\Ssp = 0.232$ at the interface.
That said, we also separately made Laplace pressure measurements of surface tension as a function of $c_\Ssp$ for a range of peak interface surfactant concentrations.
Figure~\ref{fig:GibbsMarangoni} shows that the surface tension varies linearly with the curve fit to the data  $\gamma_L(c_\Ssp) = 31.694 - 40.505\, c_\Ssp$ in dynes/cm.

As reported in a letter by Pockels to Lord Rayleigh\cite{rayleigh1891surface}, for common surfactants, such as oils and soaps, the surface tension decreases with surfactant concentration.
The significance of this observation is that an important dynamic phenomenon in surfactant-laden flows is the Marangoni effect, also known as the Gibbs–Marangoni effect, which describes the spontaneous flow of a liquid driven by gradients in surface tension along an interface. 
These gradients frequently arise from nonuniform distributions of interface
surfactant concentration, although variations in temperature or composition can also induce them. 
The underlying principle dictates that the liquid flows from regions of lower surface tension (typically areas with higher surfactant concentration) towards regions of higher surface tension (lower surfactant concentration). 
These Marangoni stresses, which originate from surfactant concentration gradients, effectively impart an apparent interfacial elasticity, influencing the flow of the bulk fluid and the interfacial stability (see Section~\ref{sec:NonequilibriumSystems}). 

\subsection{Capillary wave spectra}\label{sec:CapillaryWave}

Capillary waves due to thermal fluctuations are a mesoscopic phenomenon occurring at liquid surfaces and interfaces.
They are typically described in terms of the spectrum of mean square interface height. 
Specifically, the capillary wave spectrum is the Fourier transform of the capillary wave height of an interface. 
From the equipartition of energy at equilibrium, capillary wave theory predicts this spectrum to be\cite{Smoluchowski1908_CWT,mandelstam1913_CWT,aarts2004_CWT}
\begin{equation}
    S(k) = \langle \delta \hat{h}(k)^2 \rangle = \frac{k_B T}{L_x L_z}~\frac{1}{\gamma_\mathrm{ef}(k)\, k^2}
    \label{eq:CapWave}
\end{equation}
Although classical capillary wave theory uses a single macroscopic surface tension, studies show that at smaller length scales (larger $k$), the effective surface tension, $\gamma_\mathrm{ef}(k)$, is wavenumber dependent.\cite{mora2003x,CapWaveMD2011,Tarazona_2012} 
We write this effective surface tension as 
\begin{equation}
    \gamma_\mathrm{ef}(k) = \gamma_L ~( a_0 + a_1 \ell_\mathrm{cw}\, k + a_2 \ell_\mathrm{cw}^2\, k^2 ) 
\end{equation}
where $\ell_\mathrm{cw} = \sqrt{k_B T / \gamma_L}$ is the characteristic length scale for capillary wave fluctuations.
This wave-vector dependence incorporates effects beyond classical CWT, such as bending energy and dispersion forces. 
Our previous work\cite{MultiphaseFHD2014,BrynRP2023,bell2025comment} has shown that diffuse interface models in fluctuating hydrodynamics are in good agreement with classical capillary wave theory at low wave numbers but begin to differ at very high wavenumbers.

\begin{figure}[h]
    \centering
    \includegraphics[width=0.6\textwidth]{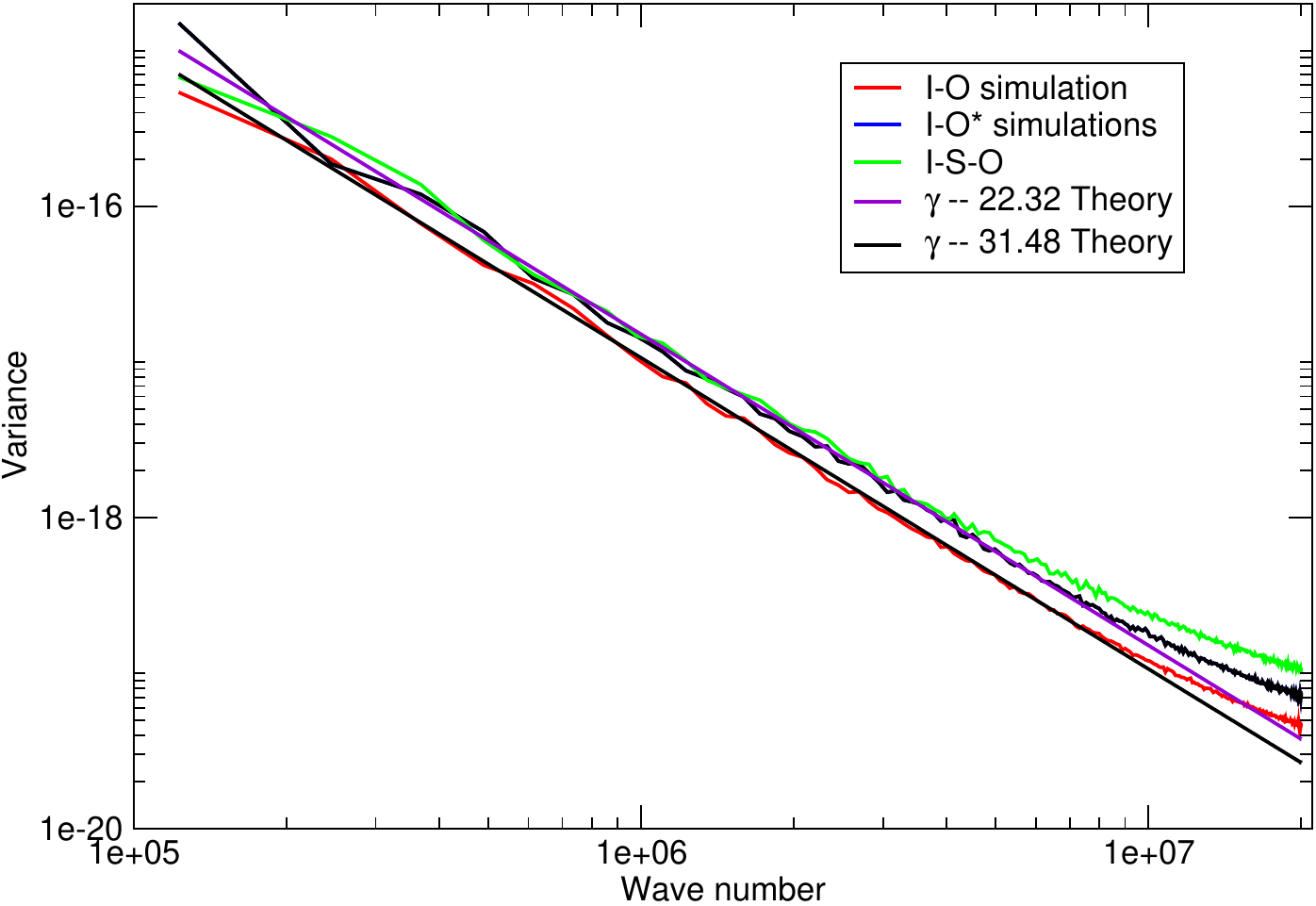}
    \caption{
    Capillary wave spectra $S(k) = \langle \delta \hat{h}(k)^2 \rangle$ versus $k$ measured in I-O, I-O*, and I-S-O simulations; theory lines given by Eq.~\ref{eq:CapWave} for constant surface tension for $\gamma_L = 31.48$ and $\gamma_L = 22.32$ dynes/cm.
    }
    \label{fig:cap_wave_spectra}
\end{figure}

\begin{figure}[h]
    \centering
    \includegraphics[width=0.90\textwidth]{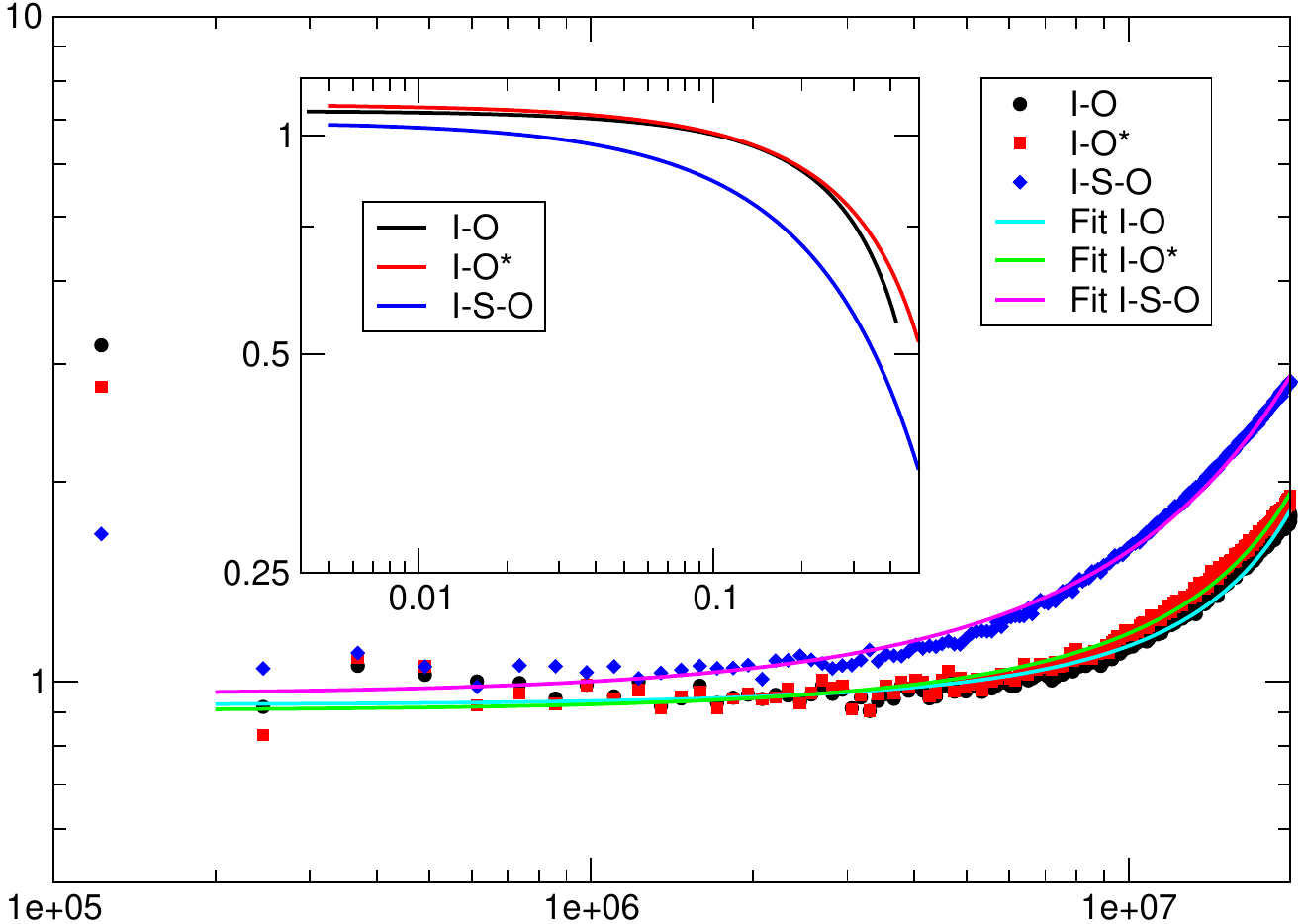}
    \caption{
    Compensated spectrum $\underline{S}(k)$ versus wave number for I-O, I-O*, and I-S-O mixtures (see legend). Points are simulation data and lines are best fits of Eq.~\ref{eq:CapWave} to the data (see Table~\ref{tab:gammaEffective} for coefficients). 
    The inset shows $\gamma_\mathrm{ef}(k)/\gamma_L$ versus $\ell_\mathrm{cw} k$ as obtained from the curve fits of $\underline{S}(k)$.
      }
    \label{fig:comp_wave_spectra}
\end{figure}

Measurements of the capillary wave spectra from our simulations for I-O, I-O*, and I-S-O mixtures are shown in Fig.~\ref{fig:cap_wave_spectra}. 
We define a dimensionless, compensated spectrum,
\begin{equation}
    \underline{S}(k) = \frac{L_x L_z \gamma_L\, k^2}{k_B T}~\langle \delta \hat{h}(k)^2 \rangle
    \label{eq:CapWaveComp}
\end{equation}
to illustrate the deviation from classical capillary wave theory.
This compensated spectrum, as measured in the simulations, is plotted in Fig.~\ref{fig:comp_wave_spectra} along with fits of Eq.~\ref{eq:CapWave} to the data. 
The inset in Fig.~\ref{fig:comp_wave_spectra} shows $\gamma_\mathrm{ef}(k)/\gamma_L$ using the best fit coefficients that are listed in Table~\ref{tab:gammaEffective}.
From this compensated spectrum it is clear that the mixture with a surfactant (I-S-O) deviates from classical capillary wave theory to a much greater extent than the two-species mixtures (I-O and I-O*). 
Furthermore, over most of the range of wave numbers considered the deviation from classical theory for I-O and I-O* is essentially the same.

\begin{table}[h]
    \centering
    \begin{tabular}{|l||l|l|l|}
        \hline
        ~ &  $a_0$& $a_1$ & $a_2$ \\
        \hline\hline
        {I-O} & 1.0828 & $-0.638087$ &  $-1.50139$ \\
        \hline
       {I-O*}  & 1.10379 & $-0.897918$ &  $-0.554654$ \\
        \hline
       {I-S-O}  & 1.04501 & $-1.88401$  & ~~0.965822 \\
        \hline
    \end{tabular}
    \caption{Best fit coefficients for $\gamma_\mathrm{ef}(k)$.  Here $\ell_{cw}^{\Isp-\Osp} = 0.2487$~nm and 
    $\ell_{cw}^{\Isp-\OAsp} =\ell_{cw}^{\Isp-\Ssp-\Osp} = 0.2094$~nm.
    }
    \label{tab:gammaEffective}
\end{table}

Finally, from observations dating back to Pliny the Elder and Benjamin Franklin, it is well known that surfactants damp capillary waves.\cite{Lohse_2023}
Undulations of the surface produce gradients of surfactant concentration and the resulting surface tension gradients induce tangential shear forces.
The flows due to these Marangoni forces result in enhanced dissipation near the interface, which is observed as damping of capillary waves in areas with surfactant (e.g., oil patch).
However, this damping cannot be measured from our capillary wave spectrum since $\langle \delta \hat{h}(k)^2 \rangle$ is a static structure factor.
This effect would be seen in the dynamic structure factor, $\langle \delta \hat{h}(k,\omega)^2 \rangle$, where $\omega$ is frequency; measurement and analysis of this damping is a topic for future work.

\section{Simulation of Non-Equilibrium Systems}\label{sec:NonequilibriumSystems}

This section presents simulation results for two non-equilibrium systems in which surfactants play a role in the dynamics. 
The first is the spread of a concentrated patch of surfactant by Marangoni convection.
The second case is the break-up of a cylinder of fluid into droplets due to the Rayleigh-Plateau instability.
Both scenarios have been studied extensively in macroscopic systems and here we consider the impact of microscopic thermal fluctuations on their dynamics.

\subsection{Spreading of a patch of surfactant}

\begin{figure}
    \centering
        \includegraphics[width=0.4\textwidth]{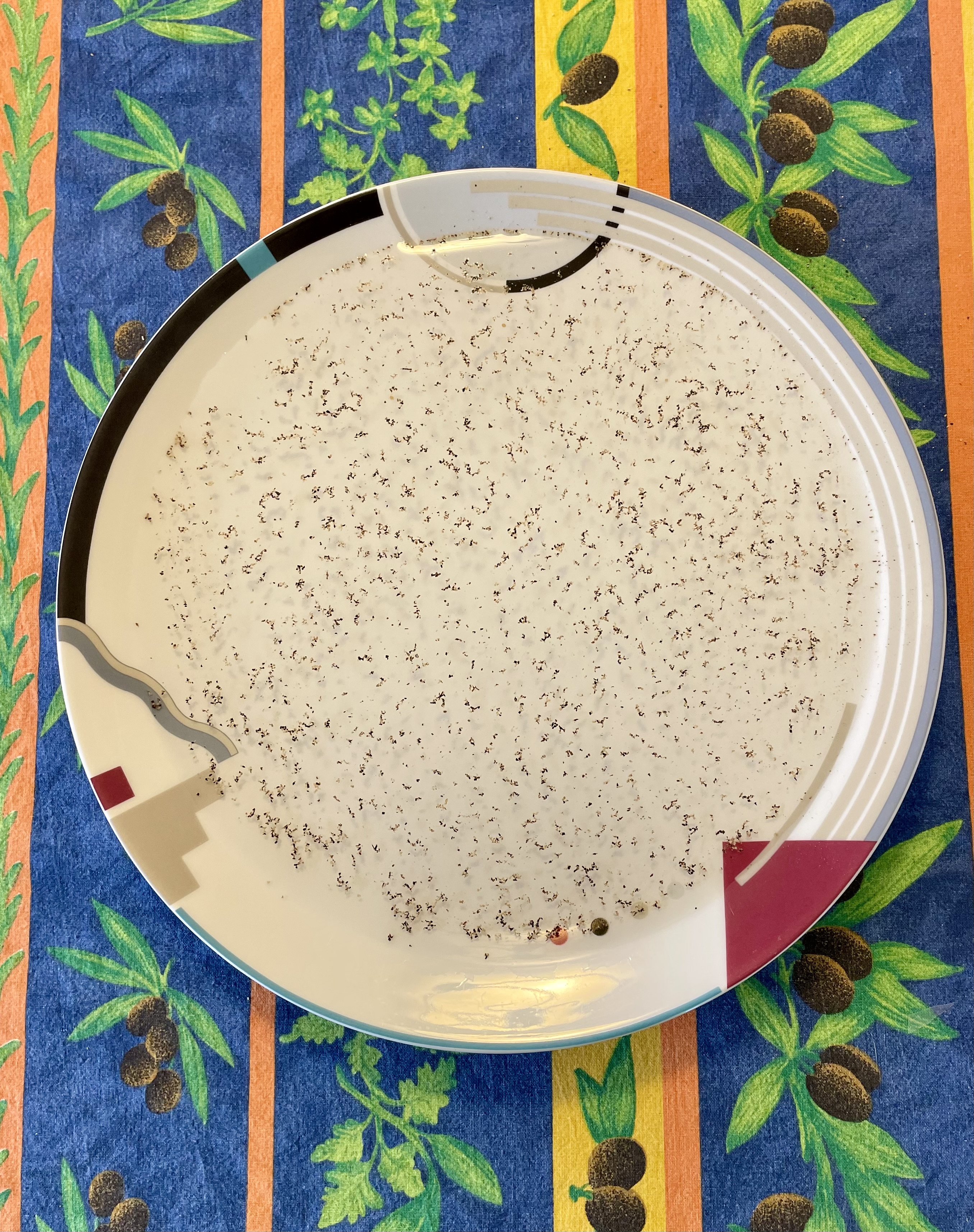}
        \includegraphics[width=0.4\textwidth]{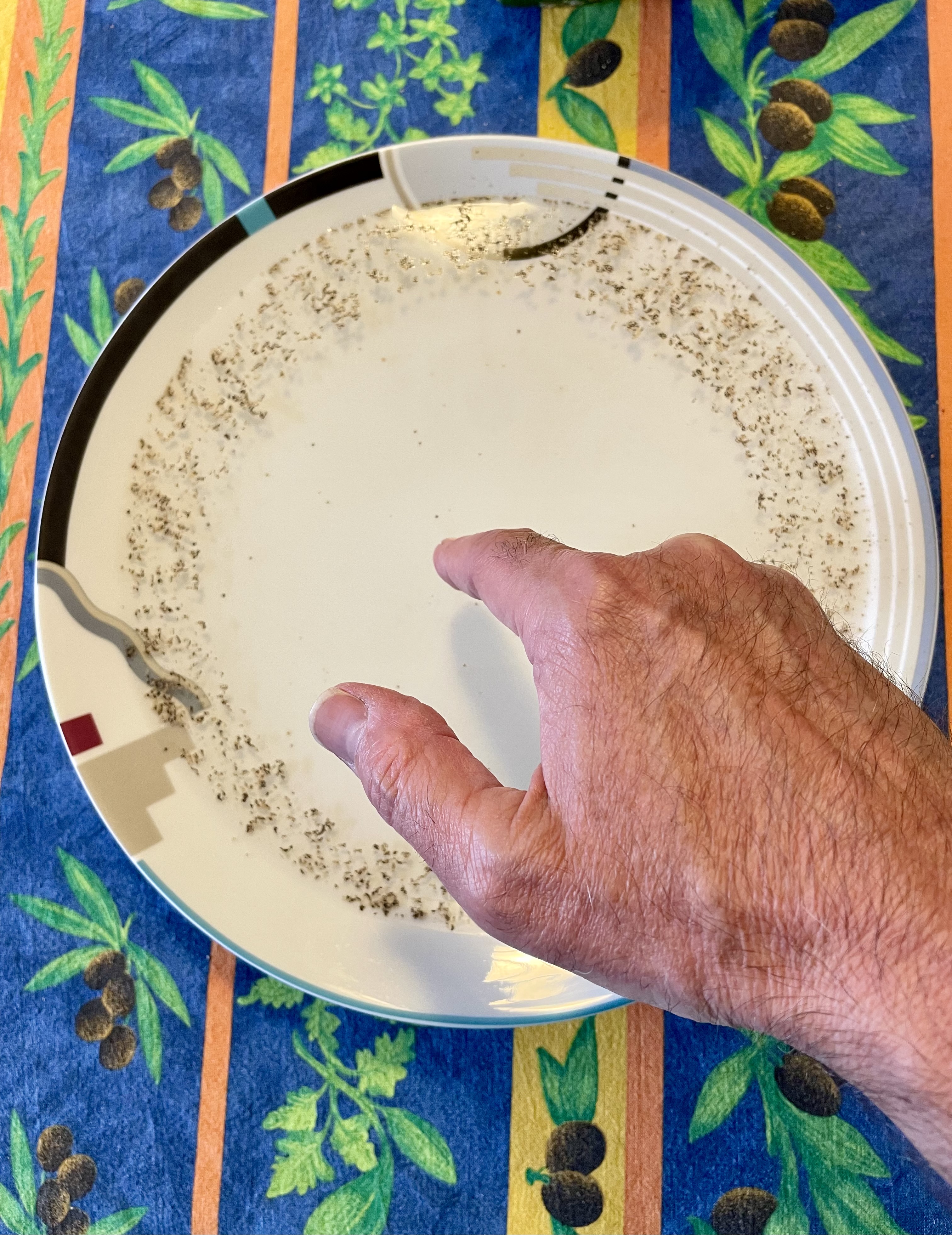}
        \caption{Demonstration of the spreading of a patch of surfactant. Left: Plate with water and a dusting of black pepper. Right: Touching the water with a finger that was dipped in dish soap.}
    \label{fig:Pepper_Soap}
\end{figure}

When a drop of surfactant is added to a liquid interface it quickly spreads across that surface.
This phenomenon can be easily demonstrated: fill a shallow plate with water, sprinkle black pepper on the surface, and then touch a soapy finger to the center of the plate (see Fig.~\ref{fig:Pepper_Soap}).
A thin soap film rapidly spreads, visibly pushing the black pepper to the edge of the plate.

This phenomenon illustrates how liquid interfaces with a non-uniform distribution of surfactant generate a dynamic driving force due to surface-tension gradients. 
These gradients induce fluid motion, known as Marangoni convection, near the surface, causing the surfactant to spread, distorting the film surface. 
The dynamics of a surfactant spreading on liquid films is of interest in a variety of industrial and biomedical applications, such as the delivery of medications or toxins via aerosol inhalation and for Surfactant Replacement Therapy (SRT)\cite{gaver1990dynamics}.

The dynamics of thin films are often modeled using lubrication theory to derive coupled evolution equations for interface height and surfactant concentration.
The solutions of these surfactant lubrication equations indicate that the film deformation differs significantly depending on the solubility of the surfactant. 
For an insoluble surfactant, the spreading often results in a shock-like structure at the leading edge of the interface, characterized by an abrupt change in film height, with corresponding thinning behind \cite{jensen1992insoluble}.  
In contrast, the spreading of soluble surfactants (especially with rapid sorption kinetics) can lead to a sharp pulse in film height just upstream of the leading edge, rather than a shock\cite{jensen1993spreading}. 
The leading-edge structure, whether it is a shock or a pulse, is also influenced and smoothed by effects such as surface diffusion and capillarity.   

\begin{figure}[h]
    \centering
     \includegraphics[width=0.85\textwidth]{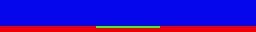} 
    \vspace{.1in}
    \includegraphics[width=0.85\textwidth]{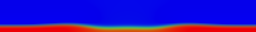} 
    \vspace{.1in}
    \includegraphics[width=0.85\textwidth]{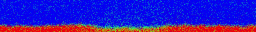}
    \caption{Snapshots of a spreading strip of surfactant on a thin (6nm) layer. From top to bottom: Initial state; Deterministic solution at $t = 40$ns; Stochastic solution from a single run at $t = 40$ns. Here we have rescaled surfactant concentration so that the surfactant distribution is easier to see.}
    \label{fig:slab_snapshots}
\end{figure}

Figure~\ref{fig:slab_snapshots} shows snapshots from our simulations for the spreading of a strip of surfactant-laden fluid initialized in the center of a system. 
In both deterministic and stochastic simulations we take a 256nm $\times$ 32nm $\times$ 1nm domain discretized with mesh spacing of 1nm in each direction.
In the upper part of the domain we set concentrations of $c_\Osp =0.92$, $c_\Isp =0.02$ and  $c_\Ssp =0.06$.
The lower part has either a thin (6nm) or a thick (8nm) layer with
$c_\Isp =0.92$, $c_\Osp =0.03$ and $c_\Ssp =0.05$ \emph{except} for a 2nm strip from 96nm $\leq x \leq$ 160nm where we increase the surfactant concentration to $c_\Ssp = 0.78$ (with $c_\Osp =0.02$ and $c_\Isp =0.20$).

\begin{figure}
    \centering
   \vspace{.1in}
    \includegraphics[width=0.4\textwidth]{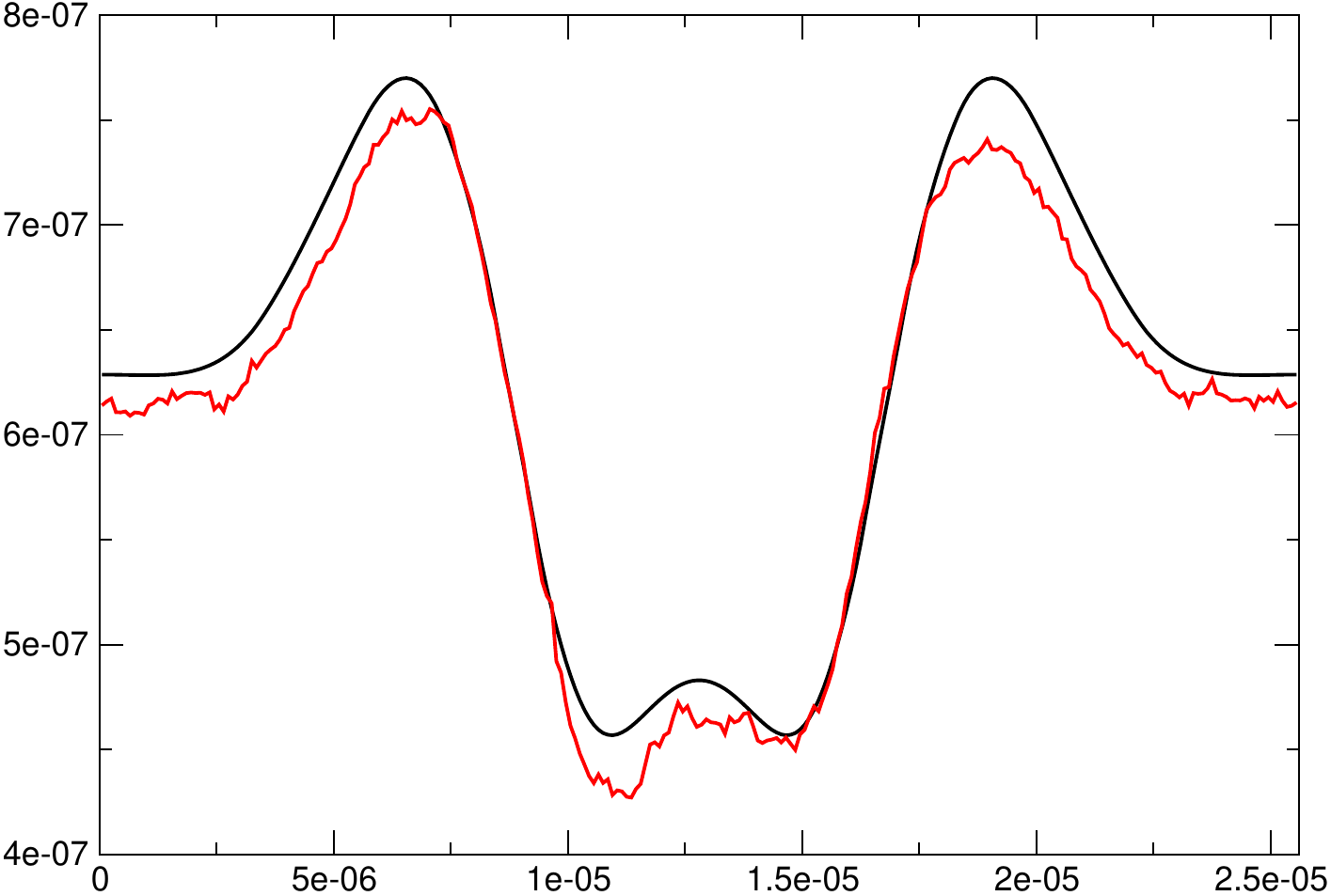}
    \includegraphics[width=0.4\textwidth]{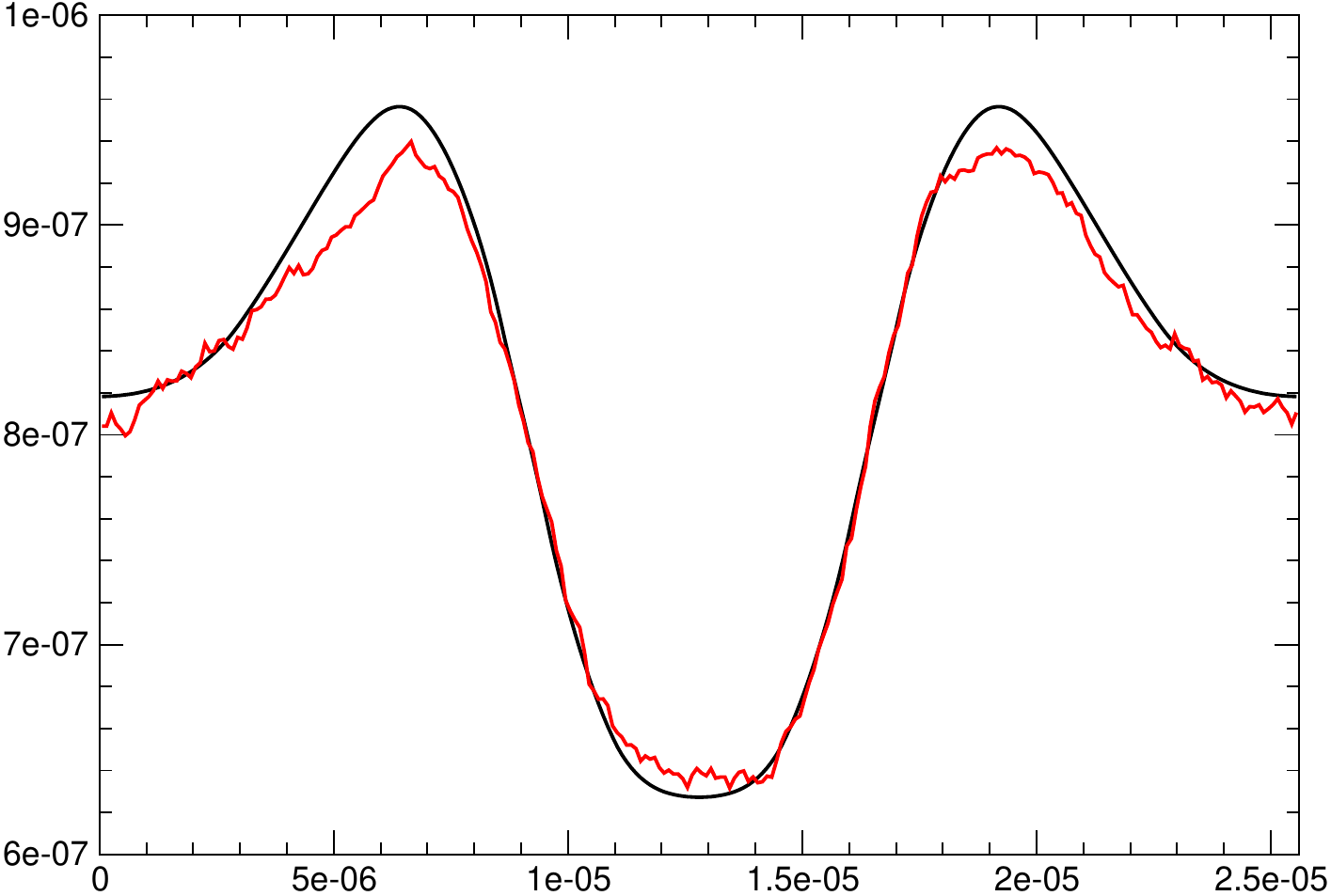}
    \caption{Interface height (in cm) versus $x$ (in cm) at $t = 40$ns for a spreading strip of surfactant on a: (left) thin (6nm) layer; (right) thick (8nm) layer.  Deterministic simulation result in black and a 100-run ensemble average of the interface height from stochastic simulations in red. 
    }
    \label{fig:slab_interface}
\end{figure}

\begin{figure}[h]
    \centering
     \includegraphics[width=0.95\textwidth]{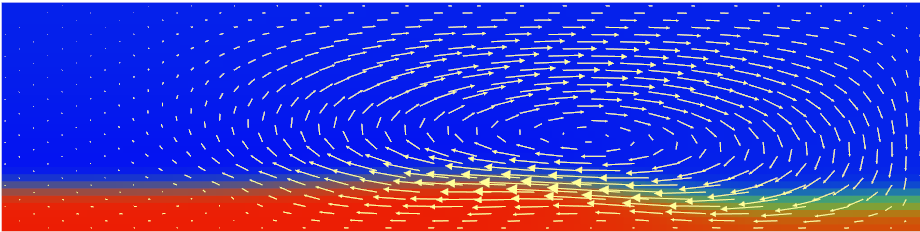} 
    \vspace{.2in}
    \includegraphics[width=0.95\textwidth]{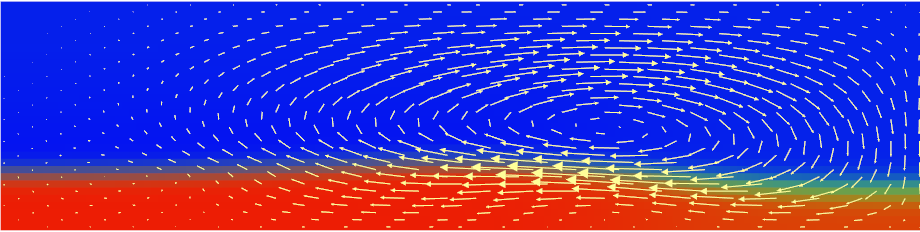}
    \caption{Fluid velocity in the left half of a deterministic simulation for a spreading strip of surfactant at $t = 40$ns: (top) 6nm layer; (bottom) 8nm layer. Arrows are scaled by the maximum fluid speed, which is 25.01 cm/s for the 6nm layer and 23.87 cm/s for the 8nm layer.}
    \label{fig:slab_velocity}
\end{figure}

From the snapshots in Fig.~\ref{fig:slab_snapshots} and the interface height profile shown in Fig.~\ref{fig:slab_interface} we see that after 40ns the surfactant strip has spread significantly. 
At $t=40$ns the peak height is located about 30nm from the initial edge of the patch; by comparison the characteristic diffusion distance is only $\sqrt{D t} \approx 8$nm.
Marangoni convection flow is seen in Figure~\ref{fig:slab_velocity}, which shows the fluid velocity for the spreading strip of surfactant for both thin (6nm) and thick (8nm) layers.

The shape of the height profiles is qualitatively similar to that obtained from numerical solutions of the surfactant lubrication equations by Jensen and Groteberg (e.g., see Fig.~2(a)\cite{jensen1992insoluble} and Fig.~1(a)\cite{jensen1993spreading}).
Finally, the height profiles in Figure~\ref{fig:slab_interface} indicate that thermal fluctuations tend to diminish the Marangoni convection as seen from the fact that the average peak heights are smaller in the stochastic case.
This is particularly true in the case of a thin layer for which the fluctuations, in some runs, resulted in the film height momentarily going to zero.

\subsection{Rayleigh-Plateau instability}

The Rayleigh-Plateau instability occurs when surface tension causes a fluid column to become unstable to small perturbations. 
The experiments of Plateau, Beer, and others in the $19\mathrm{th}$ century showed that a long cylinder (length $L$, initial radius $\InitRad$) of fluid was unstable to variations that reduced its surface area \cite{Eggers_2008,MontaneroReview2020}.
Plateau predicted that perturbations are unstable for wavelengths $\lambda \geq 2 \pi \InitRad$
and Rayleigh derived that, in the inviscid limit, the fastest growing wavelength is $\lambda_\mathrm{max} \approx 9.01 \InitRad$.
In the Stokes limit (negligible inertia), Tomotika \cite{tomotika1935instability} showed that $\lambda_\mathrm{max} \approx 11.16 \InitRad$ for a fluid cylinder immersed in a similar fluid of equal viscosity.

The stabilizing effect of a surfactant in suppressing the rupture of liquid structures, such as soap bubbles, is well established.\cite{isenberg1992science}
Qualitatively, variations in the interface area result in surfactant concentration gradients leading to surface healing due to the Gibbs-Marangoni effect when $d\gamma/dc_\Ssp < 0$.
In general, this means that the Rayleigh-Plateau instability is delayed by the presence of a surfactant.\cite{hansen1999effect, timmermans2002effect, constante2020dynamics}

At nanometer scales, thermal fluctuations play a significant role in the dynamics of the Rayleigh-Plateau instability.
For liquid-vapor systems, this mesoscopic effect has been investigated numerically with molecular dynamics\cite{Nanojets_2000,RP_MD_2006,RP_MD_2014} and theoretically using stochastic lubrication theory\cite{Nanojets_2000,Eggers_02,RayPlatFHD,Nanothreads_2020,ThreadPinch_2021}.
In earlier work \cite{BrynRP2023}, we used the numerical methods in the current paper to investigate the Rayleigh-Plateau instability in binary mixtures.
For short cylinders, with lengths less than their circumference, we found that thermal fluctuations caused these to pinch into a droplet, whereas a similar perturbed cylinder is stable deterministically.
For long cylinders, our results showed a significant effect due to thermal fluctuations in the temporal evolution of the minimum radius hastening the pinching of the cylinder into droplets.

In the present work, we expand this study to include liquid cylinders of ternary mixtures in which one species plays the role of a surfactant. 
The effect of thermal fluctuations on the Rayleigh-Plateau instability for surfactant-laden interfaces has also been investigated by molecular simulations.\cite{carnevale2024surfactant}
However, the distinct advantage of fluctuating hydrodynamics is that one can perform both deterministic and stochastic simulations to determine the specific difference made by including thermal fluctuations.
Furthermore, FHD advances with timesteps on the order of a picosecond as compared with molecular dynamics timesteps which are typically on the order of a femtosecond.

As described in Section~\ref{sec:MixtureCompositions}, we separately consider three mixture cases: I-O, I-O*, and I-S-O mixtures.
The cylindrical initial condition was produced by replicating the corresponding 2D droplet steady state (see Section~\ref{sec:LaplacePressure}) in the $z$ direction; boundaries are periodic in all directions.
For the deterministic simulations we perturb the system by advancing the simulation for 5000 steps (1 nanosecond) with thermal fluctuations and then continue the simulation with fluctuations turned off by setting the stochastic flux terms to zero.

There are several dimensionless numbers that characterize the dynamics of the Rayleigh-Plateau instability. The Ohnesorge number, $\OhNum = \eta/\sqrt{\rho\InitRad\gamma}$, compares viscous forces to inertial and surface tension forces, characterizing the relative importance of shear viscosity, $\eta$, to surface tension, $\gamma$.
In our simulations of cylinders with an initial radius of 8nm,  $\OhNum \approx 2.0$ for the I-O case and $\OhNum \approx 2.4$ for the I-O* and I-S-O cases.

In the stochastic simulations, to compute a robust effective radius in the presence of fluctuations we use the procedure developed in Barker \textit{et al.}\cite{BrynRP2023}.  
First, a filter is applied to the data, defining
\begin{align}
    \tilde{C} = \max \left( \min \left( \frac{ C + \Delta C_f - 1/2}{2\Delta C_f} , 1 \right), 0\right)
\end{align}
where $C = c_\Isp +c_\Ssp$. Note that by filtering based on the combined concentration of inner fluid and surfactant, we avoid incorrectly identifying the case of a cylinder with a narrow surfactant bridge as a pinched droplet.
This filter modifies the thickness of an interface, specifically $\tilde{\ell} = 2 \Delta C_f \ell$; for our calculations we chose $\Delta C_f = 0.1$ so the filter reduces the interface thickness by a factor of five. 
The effective radius for a slice of cells is then computed as
\begin{align}
    R(x,t)=\sqrt{\frac{1}{\pi}(\Delta z)(\Delta y)\sum_{i=1}^{N_x}\sum_{j=1}^{N_y}
    \tilde{C}_{i,j,k}^n},
\end{align} 
where $x = i \Delta x$ and $t = n \Delta t$.  
The minimum cylinder radius is $R_\mathrm{min}(t) = \min_x R(x,t)$.

\subsubsection{Short cylinders}

\begin{figure}[h]
    \centering
    \includegraphics[width=0.24\textwidth]{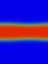}
    \includegraphics[width=0.24\textwidth]{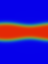}
    \includegraphics[width=0.24\textwidth]{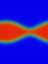}
    \includegraphics[width=0.24\textwidth]{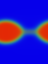}
    \caption{Snapshots from a deterministic simulation of a short I-S-O cylinder  at $t = 100$, 150, 180, 183.4ns. The color map is: Species I (red); Species S (green); Species O (blue).
    }
    \label{fig:det_surf_stub}
\end{figure}

\begin{figure}[h]
    \centering
    \includegraphics[width=0.8\textwidth]{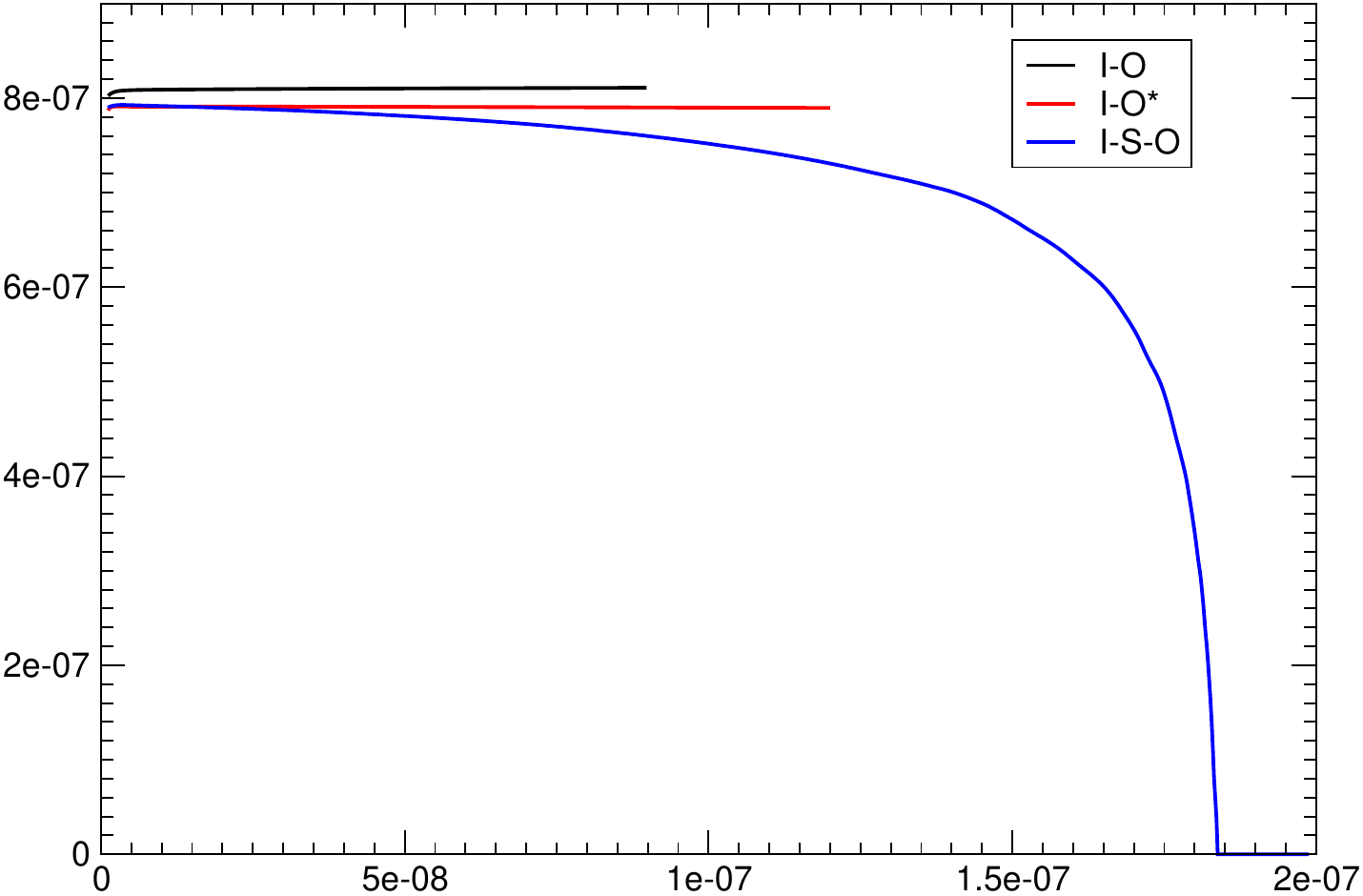}
    \caption{Minimum radius $R_\mathrm{min}(t)$ (in cm) versus time (in s) for deterministic runs of short cylinders of I-O (black), I-O* (red), and I-S-O mixtures (blue).}
\label{fig:det_hist}
\end{figure}

First, we consider deterministic simulations of short ($L = 48$nm) cylinders in a 64nm $\times$ 64nm $\times$ 48nm domain.
The initial radius was $\InitRad \approx$ 8nm so they are expected to be marginally stable since $\mathcal{A} = L / 2\pi \InitRad \approx 0.96 \lessapprox 1$.
Note that this result is independent of the value of the surface tension.
Figure~\ref{fig:det_surf_stub} shows snapshots from a deterministic simulation of a short I-S-O cylinder and, interestingly, here we see pinching for the cylinder with surfactant.
The minimum radius $R_\mathrm{min}(t)$ as a function of time for all three cases is shown in Figure~\ref{fig:det_hist}. 
As in our previous work~\cite{BrynRP2023}, we found that for binary (I-O or I-O*) mixtures these short cylinders are stable against the Rayleigh-Plateau instability.

\begin{figure}
    \centering
    \includegraphics[width=0.3\textwidth]{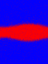}
    \includegraphics[width=0.3\textwidth]{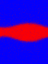}
    \includegraphics[width=0.3\textwidth]{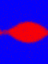}

    \includegraphics[width=0.3\textwidth]{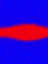}
    \includegraphics[width=0.3\textwidth]{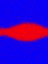}
    \includegraphics[width=0.3\textwidth]{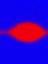}

    \includegraphics[width=0.3\textwidth]{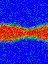}
    \includegraphics[width=0.3\textwidth]{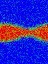}
    \includegraphics[width=0.3\textwidth]{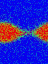}
      
    \caption{Snapshots from stochastic simulations of short cylinders. Top to bottom: I-O, I-O*, I-S-O; left to right at different times (top: 70ns, 75ns, 80ns; middle: 75ns, 80ns, 85ns; bottom: 70ns, 75ns, 80ns) }
    \label{fig:short}
\end{figure}

\begin{figure}
    \centering
    \includegraphics[width=0.3\textwidth]{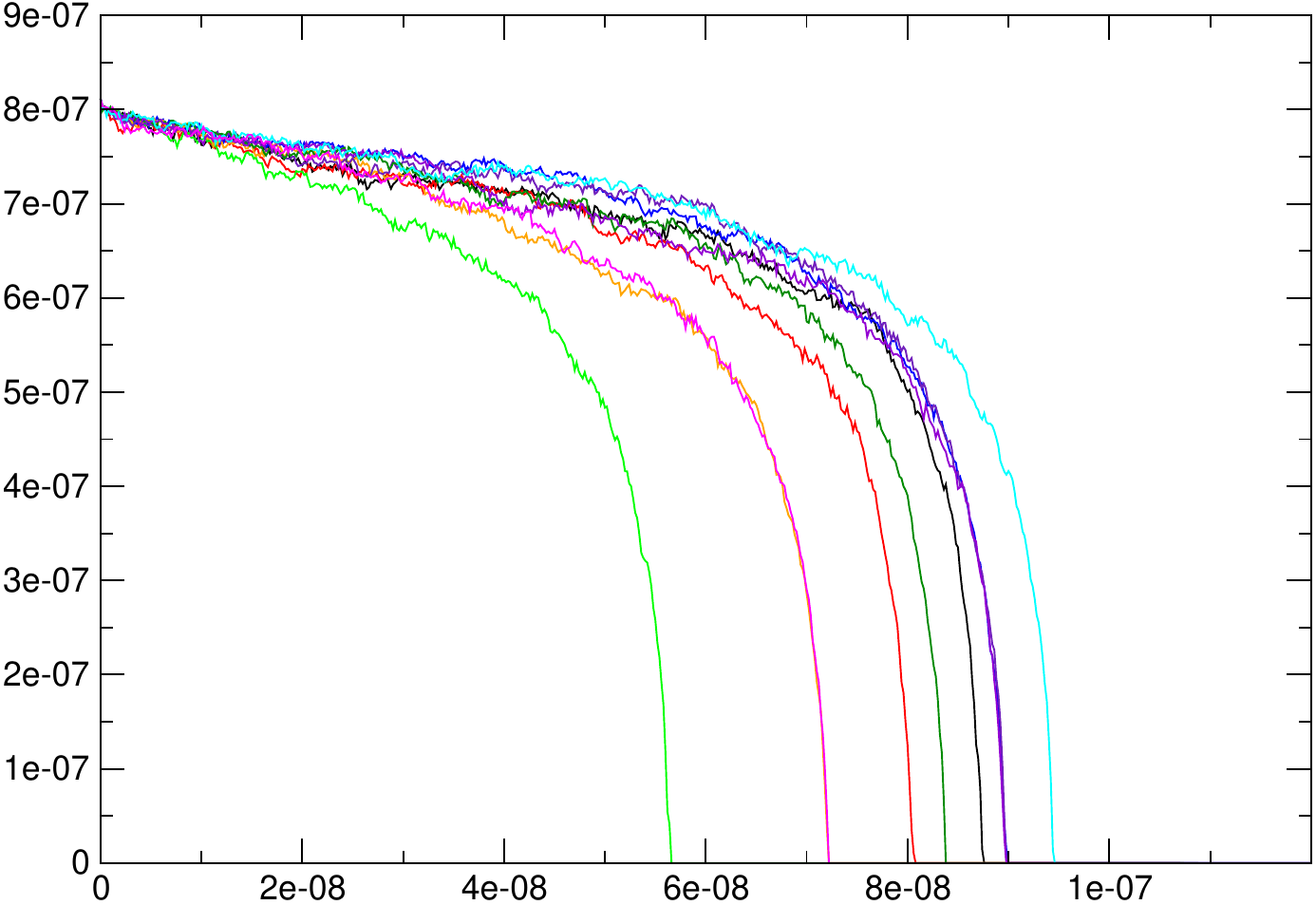}
    \includegraphics[width=0.3\textwidth]{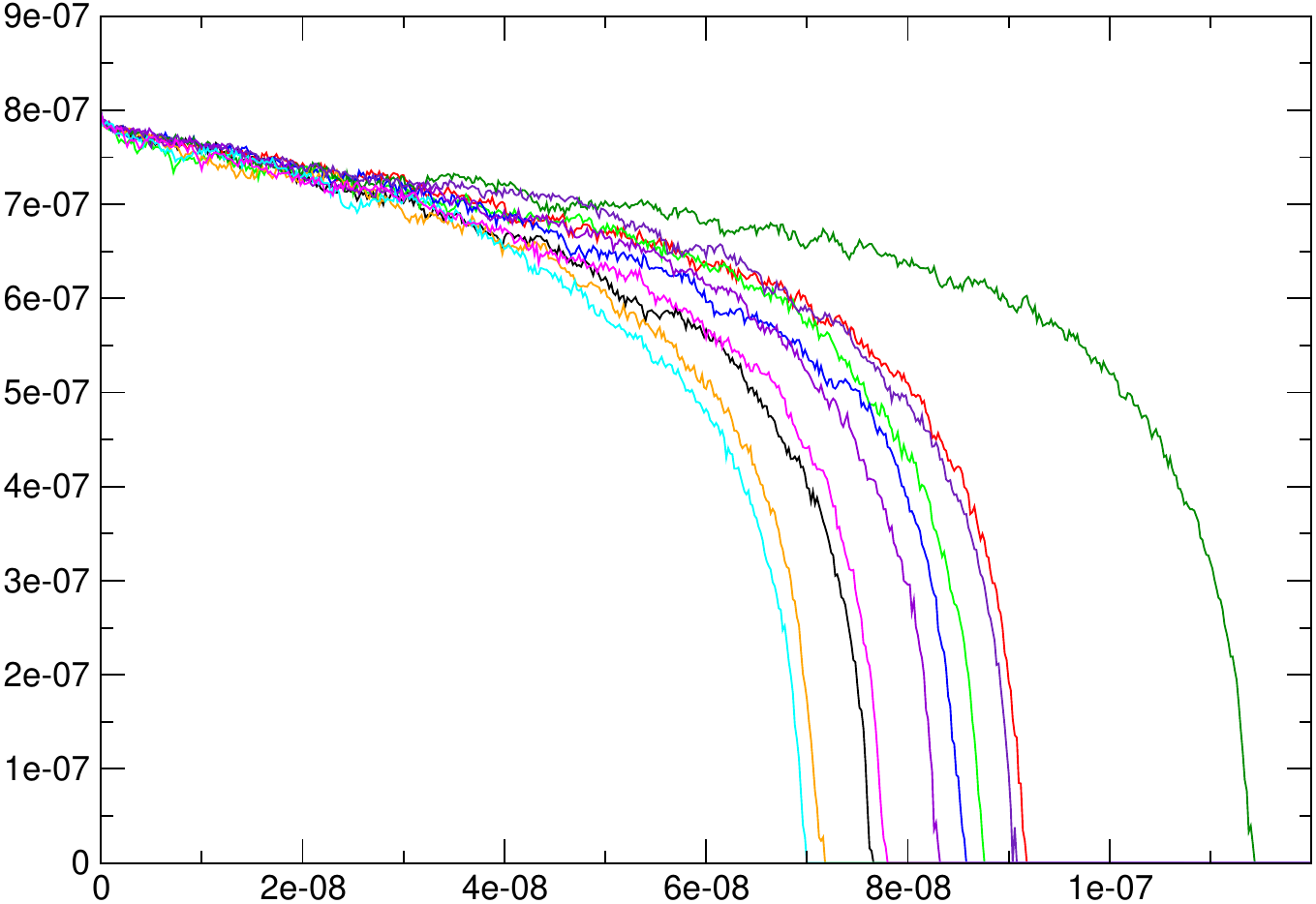}
    \includegraphics[width=0.3\textwidth]{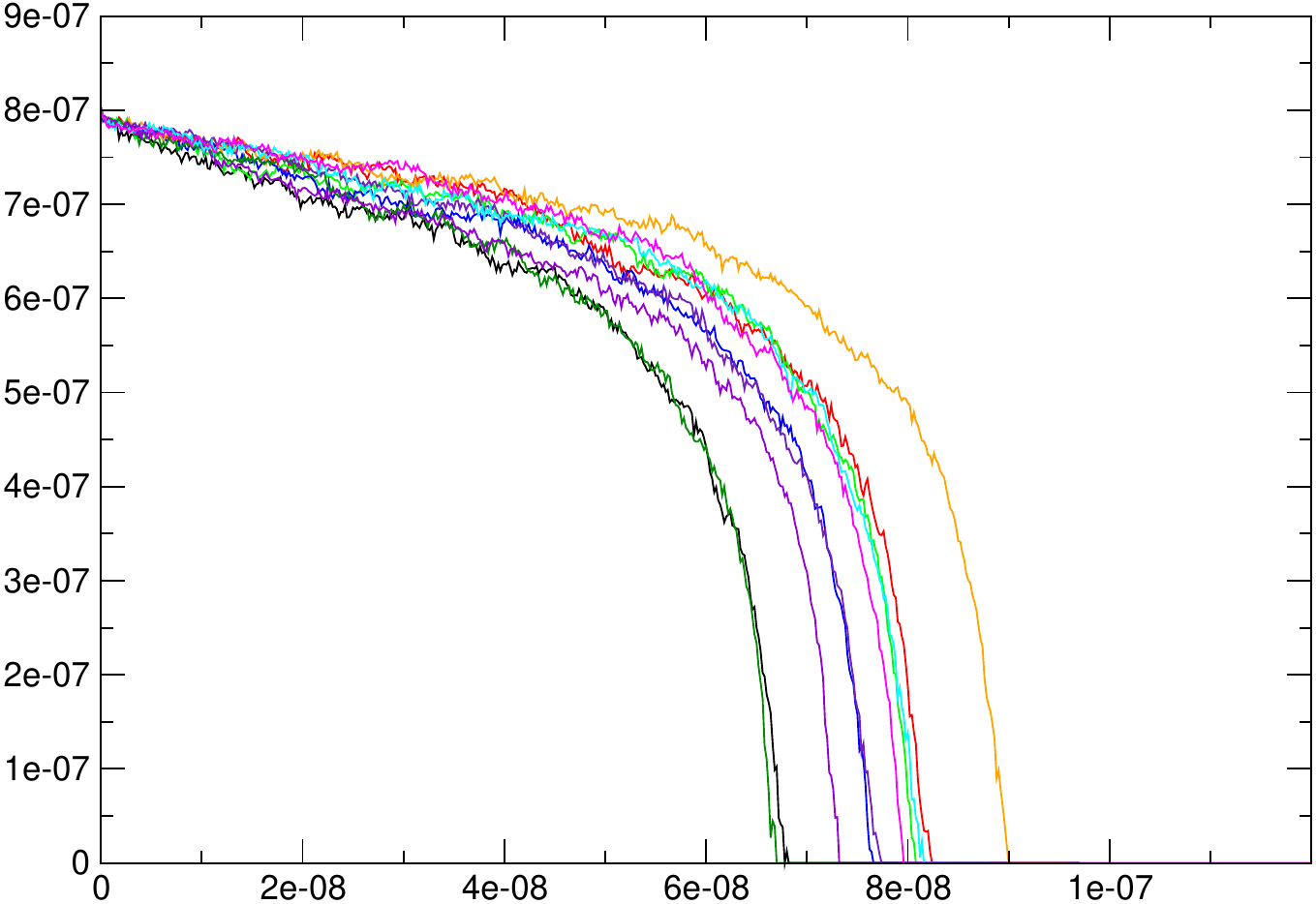}

    \caption{Minimum cylinder radius as a function of time for a 10 run ensemble of short cylinders. Left to right: I-O, I-O*, I-S-O. Average times for pinching are $\langle\check{t}\rangle = 80.31$ns, 84.21ns and 78.27ns with standard deviations of 11.99ns, 11.20ns, and 7.24ns, respectively.}
    \label{fig:short_minrad}
\end{figure}

\begin{figure}
    \centering
    \includegraphics[width=0.3\textwidth]{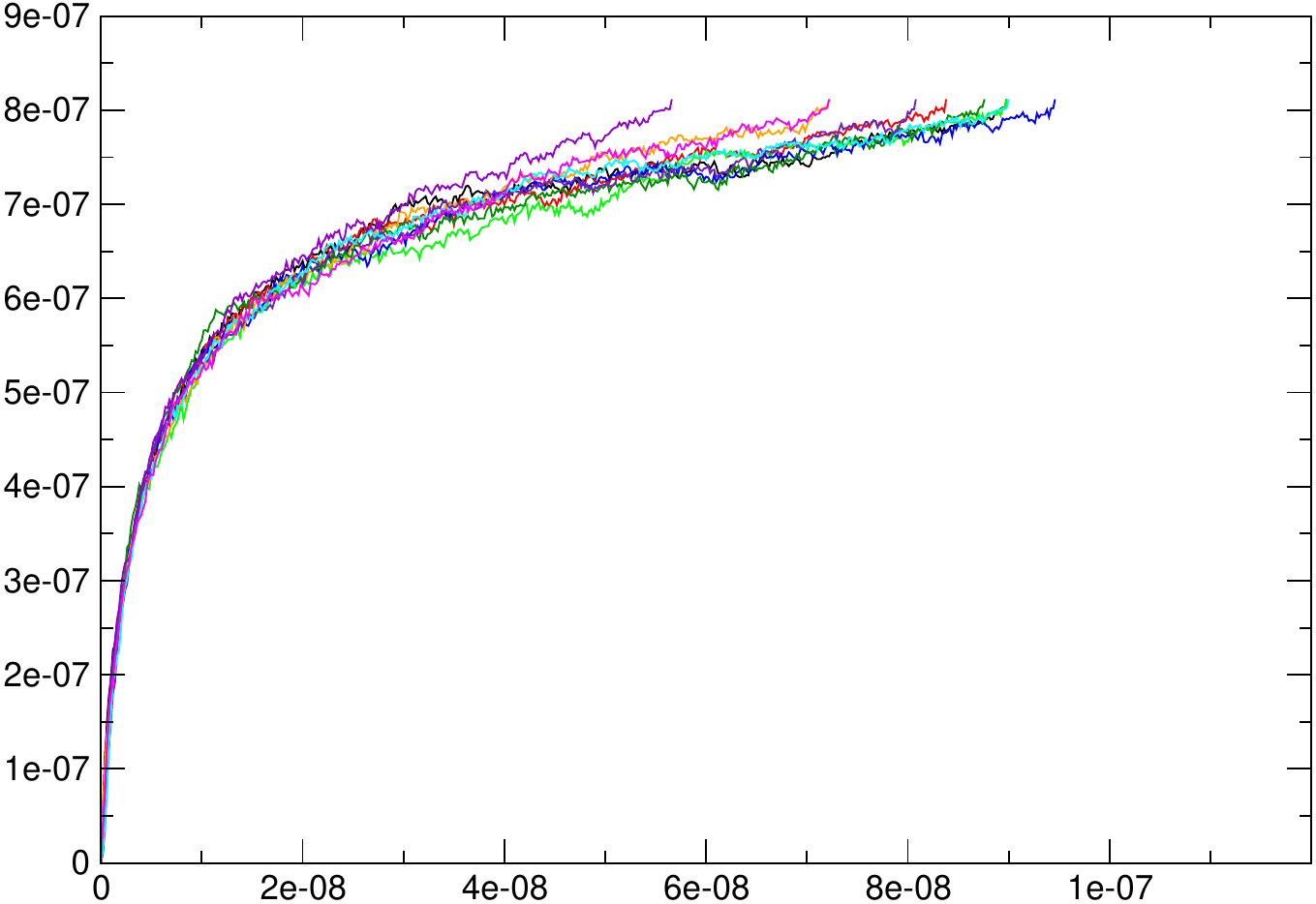}
    \includegraphics[width=0.3\textwidth]{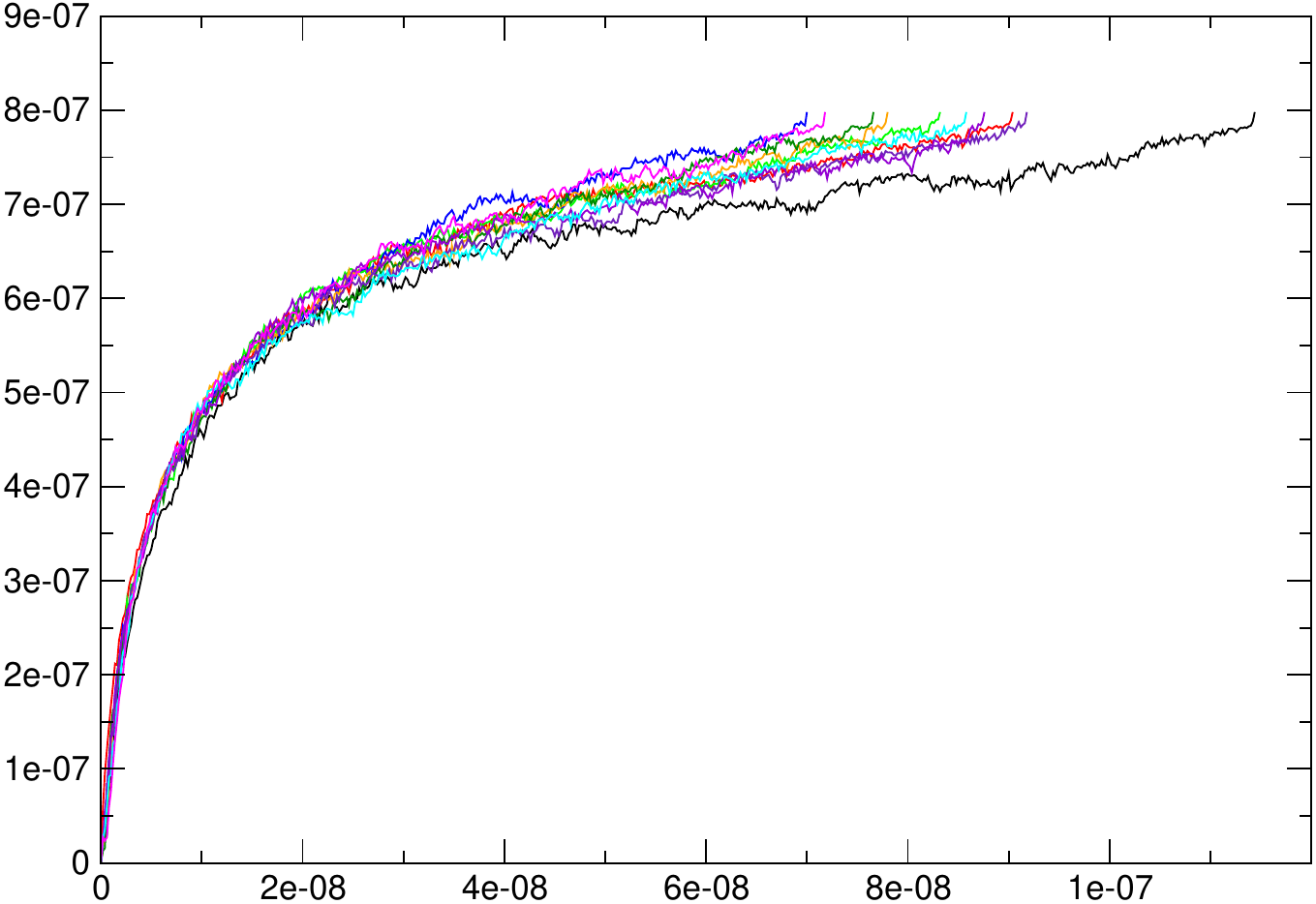}
    \includegraphics[width=0.3\textwidth]{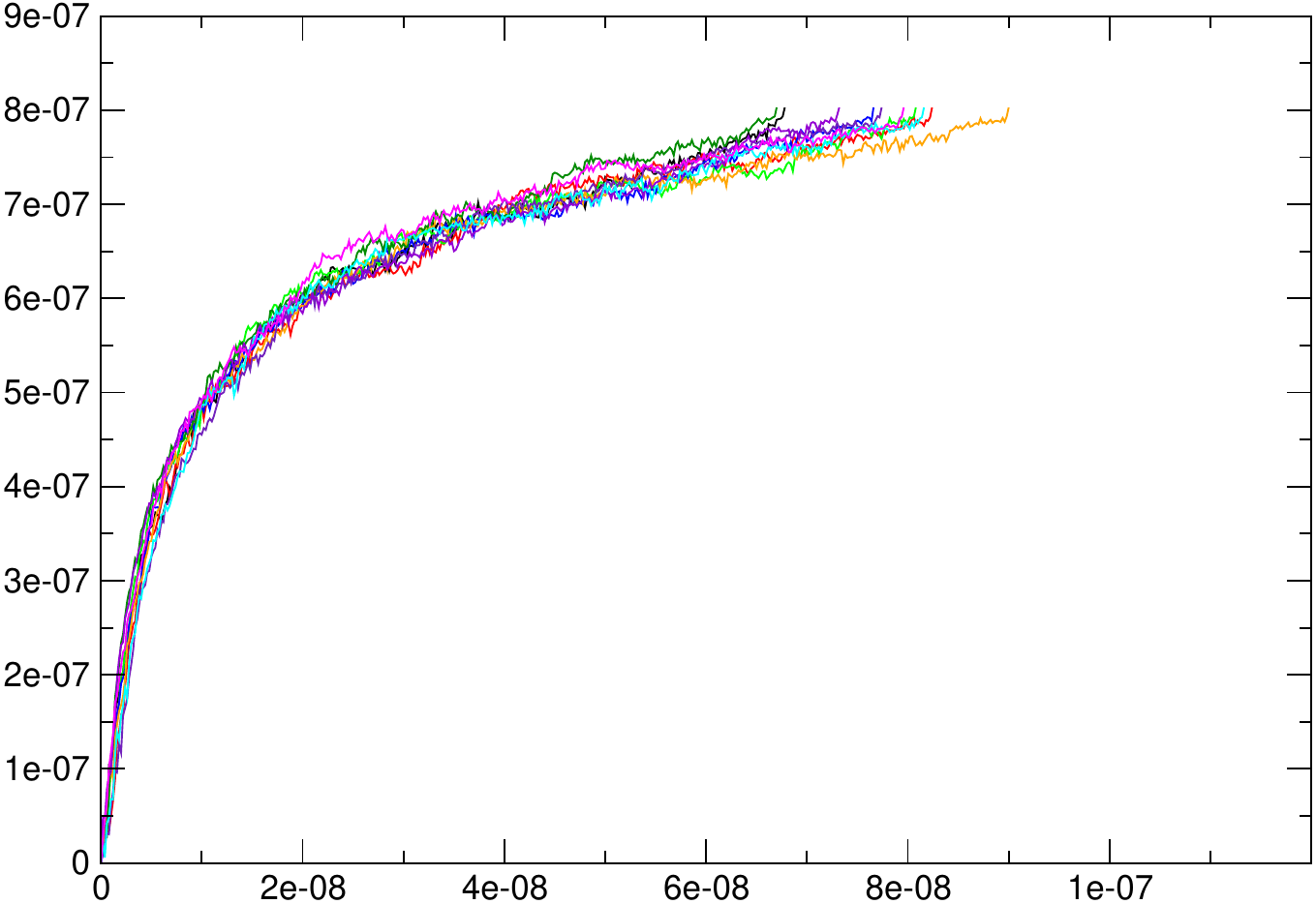}

    \caption{Mininum cylinder radius, $R_\mathrm{min}(t)$, versus $\check{t} - t$ (time before pinch). Left to right: I-O, I-O*, I-S-O; also see Fig.~\ref{fig:short_minrad}.}
    \label{fig:short_minradrev}
\end{figure}

\begin{figure}
    \centering
    \includegraphics[width=0.9\textwidth]{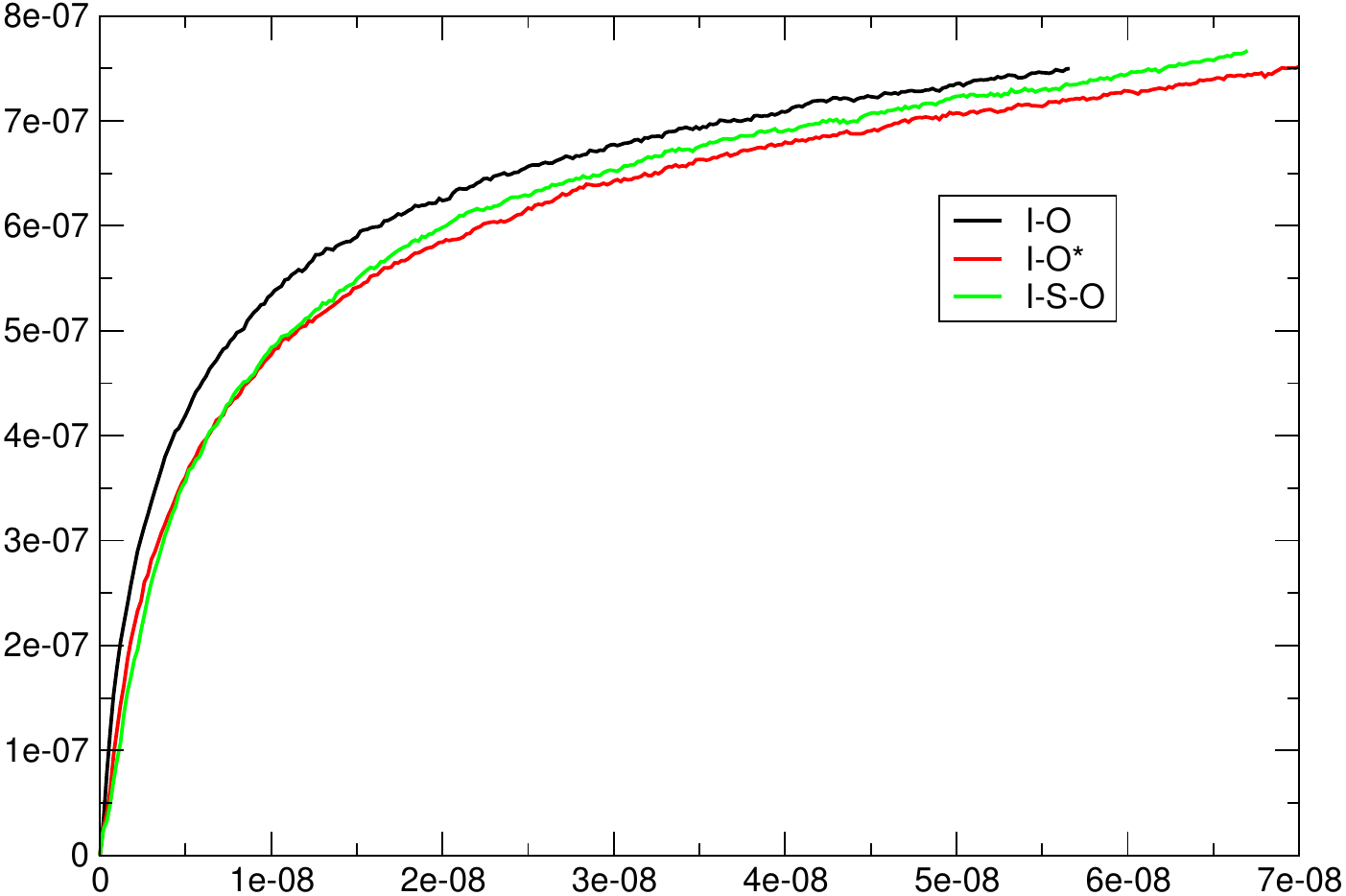}

    \caption{Ensemble average from short cylinder runs of the mininum cylinder radius versus $\check{t} - t$ (time before pinch) for I-O (black), I-O* (red), and I-S-O (green) mixtures. 
    }
    \label{fig:short_avgminradrev}
\end{figure}

Next we turn to the stochastic simulations of these short cylinders. Snapshots from a typical run for I-O, I-O*, and I-S-O mixtures are shown in Fig.~\ref{fig:short}.
In this set of simulations, we observe that the I-O* mixture pinches into a droplet a few nanoseconds later than the other two cases.
To obtain a more detailed comparison of the different mixtures we performed additional simulations to create an ensemble of 10 simulations for each mixture. 
For each mixture we plot in Fig.~\ref{fig:short_minrad} the minimum cylinder radius versus time for each simulation in the ensemble.
This figure shows that $\check{t}$, the ``pinching time'' (earliest time at which $R_\mathrm{min}(t) = 0$), varies significantly from run to run. 
The average pinching times in this 10 run ensemble are $\langle\check{t}\rangle = 77.44$ns, 82.72ns and 79.52ns for I-O, I-O*, and I-S-O mixtures, respectively; 
standard deviations are 13.55ns, 8.25ns, and 8.15ns.
Plotting $R_\mathrm{min}(t)$ versus $\check{t} - t$ shows that for each mixture the curves have a similar form as shown in Fig..~\ref{fig:short_minradrev}.
Furthermore comparing the average of these curves for each mixture in Fig.~\ref{fig:short_avgminradrev} shows that the surfactant has minimal effect.
Note that for short cylinders the results from both deterministic and stochastic runs were \emph{not} well fit by a power law of the form $R_\mathrm{min}(t) \propto (\check{t} - t)^a$.
This is in contrast with the results for long cylinders ($L \gg 2\pi \InitRad$), as shown in the next section.

\begin{figure}
    \centering
    \includegraphics[width=0.3\textwidth]{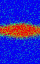}
    \includegraphics[width=0.3\textwidth]{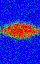}
    \includegraphics[width=0.3\textwidth]{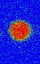}
      
    \caption{
    Snapshots from a stochastic simulation of a very short ($\mathcal{A}\approx 0.8$) I-S-O cylinder at $t = 150$, 160, 170ns. 
    }
    \label{fig:vstub}
\end{figure}

Finally, Figure~\ref{fig:vstub} shows snapshots from a stochastic simulation of a very short ($L = 40$nm) cylinder with surfactant (I-S-O mixture) in a simulation domain of $64 \times 64 \times 40$ cells.
As we see, with thermal fluctuations this very short cylinder ($\mathcal{A} \approx 0.8$) pinches into a droplet, as did very short cylinders with binary mixtures, either I-O or I-O*. 
By comparison, for all mixture cases deterministic simulations (not shown) of very short cylinders were stable in the absence of thermal fluctuations.

\subsubsection{Long cylinders}

\begin{figure}[h]
    \centering
    \includegraphics[width=0.3\textwidth]{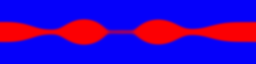}
   \includegraphics[width=0.3\textwidth]{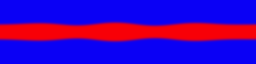}
   \includegraphics[width=0.3\textwidth]{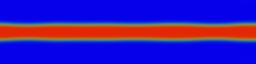}

       \includegraphics[width=0.3\textwidth]{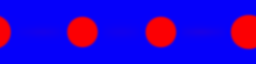}
   \includegraphics[width=0.3\textwidth]{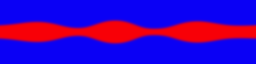}
   \includegraphics[width=0.3\textwidth]{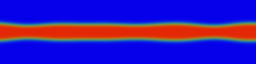}

       \includegraphics[width=0.3  \textwidth]{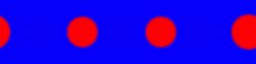}
   \includegraphics[width=0.3\textwidth]{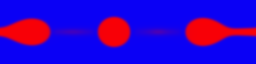}
   \includegraphics[width=0.3\textwidth]{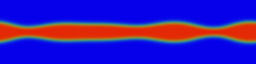}
   
       \includegraphics[width=0.3\textwidth]{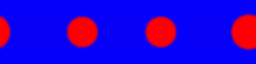}
   \includegraphics[width=0.3\textwidth]{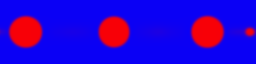}
   \includegraphics[width=0.3\textwidth]{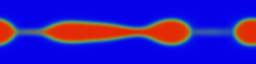}
    \caption{Snapshots from deterministic runs for long ($L = 256$nm) cylinders taken at $t = 40$, 50, 60, 70 ns (top to bottom) for I-O, I-O*, I-S-O mixtures (left to right).}
    \label{fig:det_long}
\end{figure}

In the short-cylinder scenarios described above, an unstable cylinder pinches into a single droplet.
In contrast, we see from Figure~\ref{fig:det_long} that in deterministic simulations of long cylinders ($L \gg 2\pi \InitRad$) the Rayleigh-Plateau instability leads to multiple droplets.
Comparing the two non-surfactant cases, we see that the I-O case, which has a surface tension higher than that of I-O*, forms droplets sooner. 
Although the I-O* and I-S-O cases have similar surface tensions, the Rayleigh-Plateau instability is delayed in the case with surfactant.

\begin{figure}[h]
    \centering
    \includegraphics[width=0.45\textwidth]{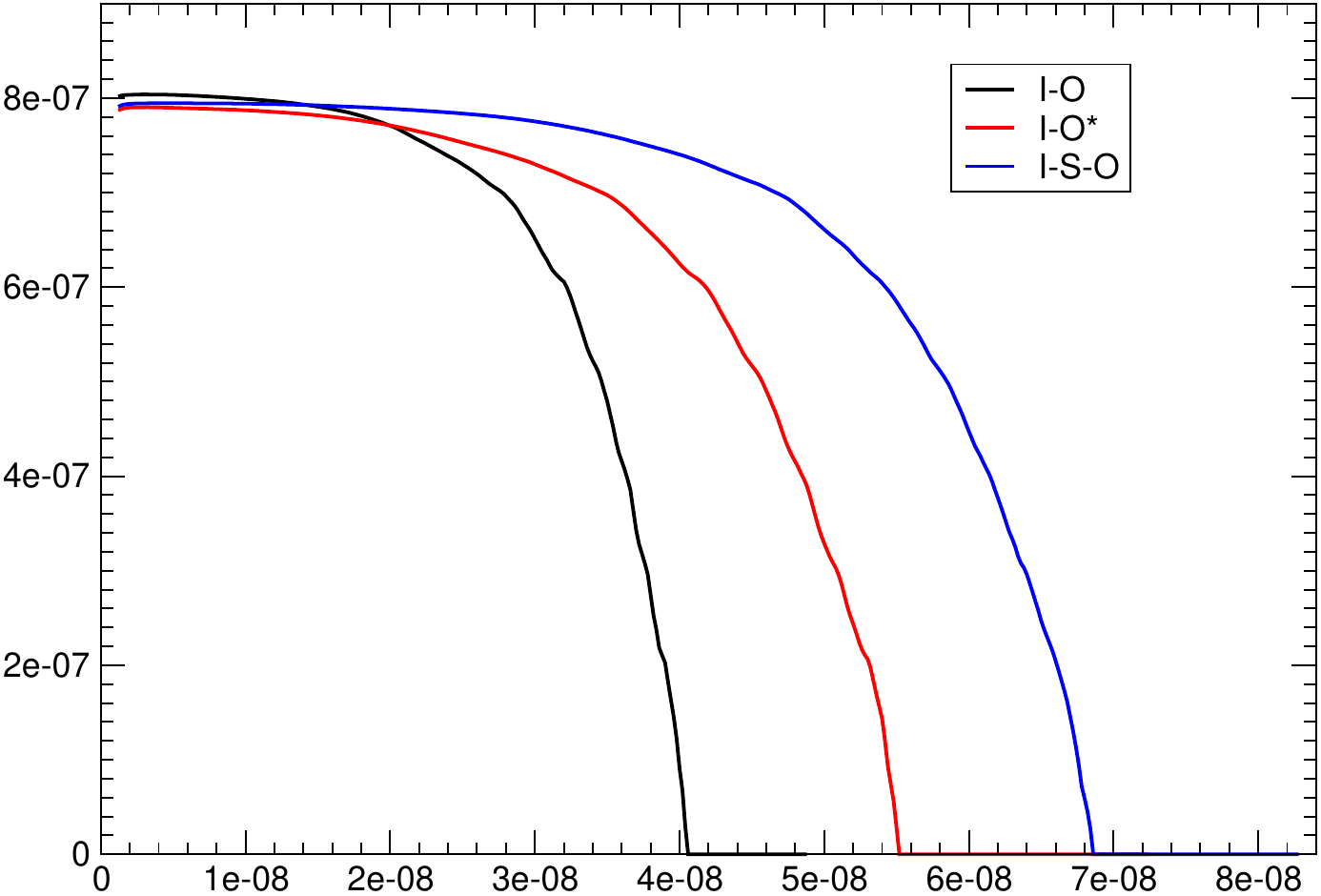}
    \includegraphics[width=0.45\textwidth]{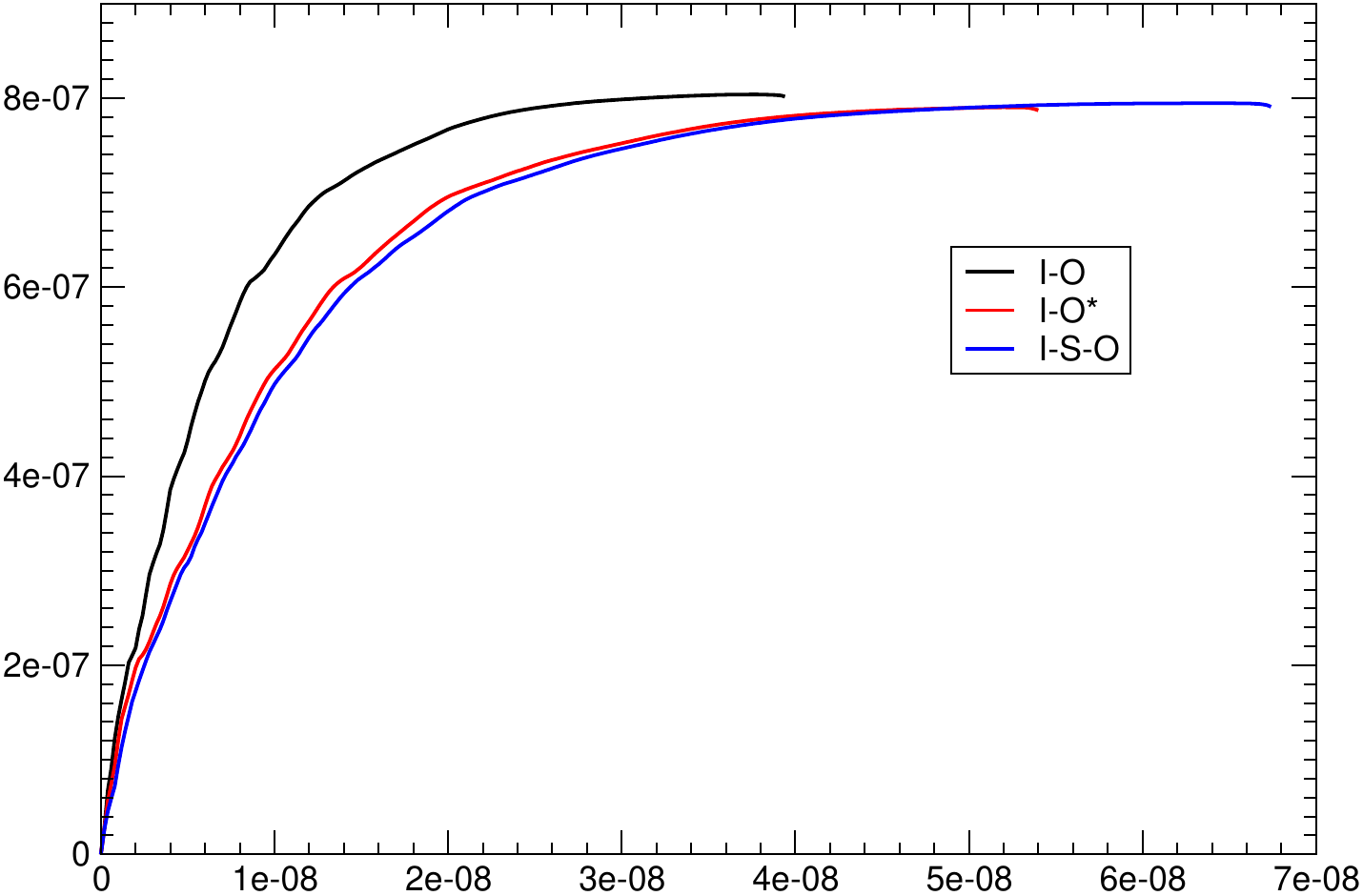}
    
    \caption{Minimum radius $R_\mathrm{min}(t)$ versus $t$ (left) and versus $\check{t}-t$ (right) from deterministic simulations of long cylinders. 
    }
    \label{fig:det_long_rmin}
\end{figure}

Figure~\ref{fig:det_long_rmin} shows the minimum radius as a function of time from deterministic simulations of long cylinders.
As expected, the pinching time, $\check{t}$, is longer for the I-O* case than for the I-O case, since the larger surface tension accelerates the instability.
Furthermore, $\check{t}$ is longer still for the I-S-O case since the Marangoni stress, produced by the surfactant gradient in the narrowing neck of the cylinder, further delays the pinching of the cylinder.
Similar results are observed in experiments, for example see Figure~6 in Kovalchuk \textit{et al.}\cite{kovalchuk2017kinetics}.
We find that the minimum radius is reasonably approximated as a power law of the form $R_\mathrm{min}(t) \propto (\check{t} - t)^{\check{a}}$.
From a fit of $R_\mathrm{min}$ between 1nm and 5nm we get: for I-O, $\check{a} = 0.7015$; for I-O*, $\check{a} = 0.6105$; and for I-S-O, $\check{a} = 0.6711$.
These results are consistent with the theoretical prediction\cite{keller1983surface} of $\check{a} = 2/3$ in the inertial regime (small $\mathrm{Oh}$) and with experimental measurements of $\check{a} \approx 2/3$ for liquid cylinders with soluble surfactants\cite{kovalchuk2018effect}.

\begin{figure}
    \centering
    \includegraphics[width=0.3\textwidth]{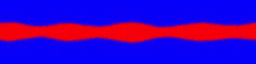}
   \includegraphics[width=0.3\textwidth]{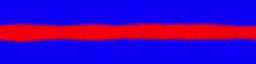}
   \includegraphics[width=0.3\textwidth]{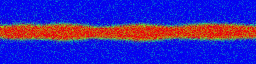}

    \includegraphics[width=0.3\textwidth]{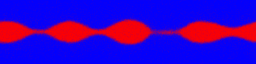}
   \includegraphics[width=0.3\textwidth]{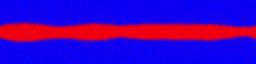}
   \includegraphics[width=0.3\textwidth]{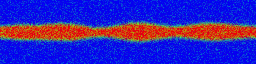}
   
       \includegraphics[width=0.3\textwidth]{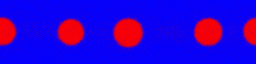}
   \includegraphics[width=0.3\textwidth]{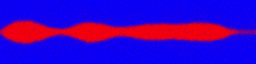}
   \includegraphics[width=0.3\textwidth]{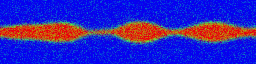}
   
       \includegraphics[width=0.3\textwidth]{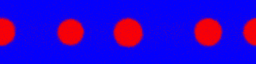}
   \includegraphics[width=0.3\textwidth]{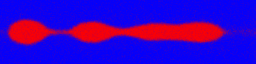}
   \includegraphics[width=0.3\textwidth]{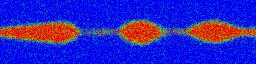}
    \caption{Snapshots at $t = 30$, 37, 46, and 49ns (top to bottom) from stochastic simulations of long ($L = 256$nm) cylinders for (left to right) I-O, I-O*, and I-S-O mixtures.}
    \label{fig:stoch_long}
\end{figure}

\begin{figure}
    \centering
    \includegraphics[width=0.3\textwidth]{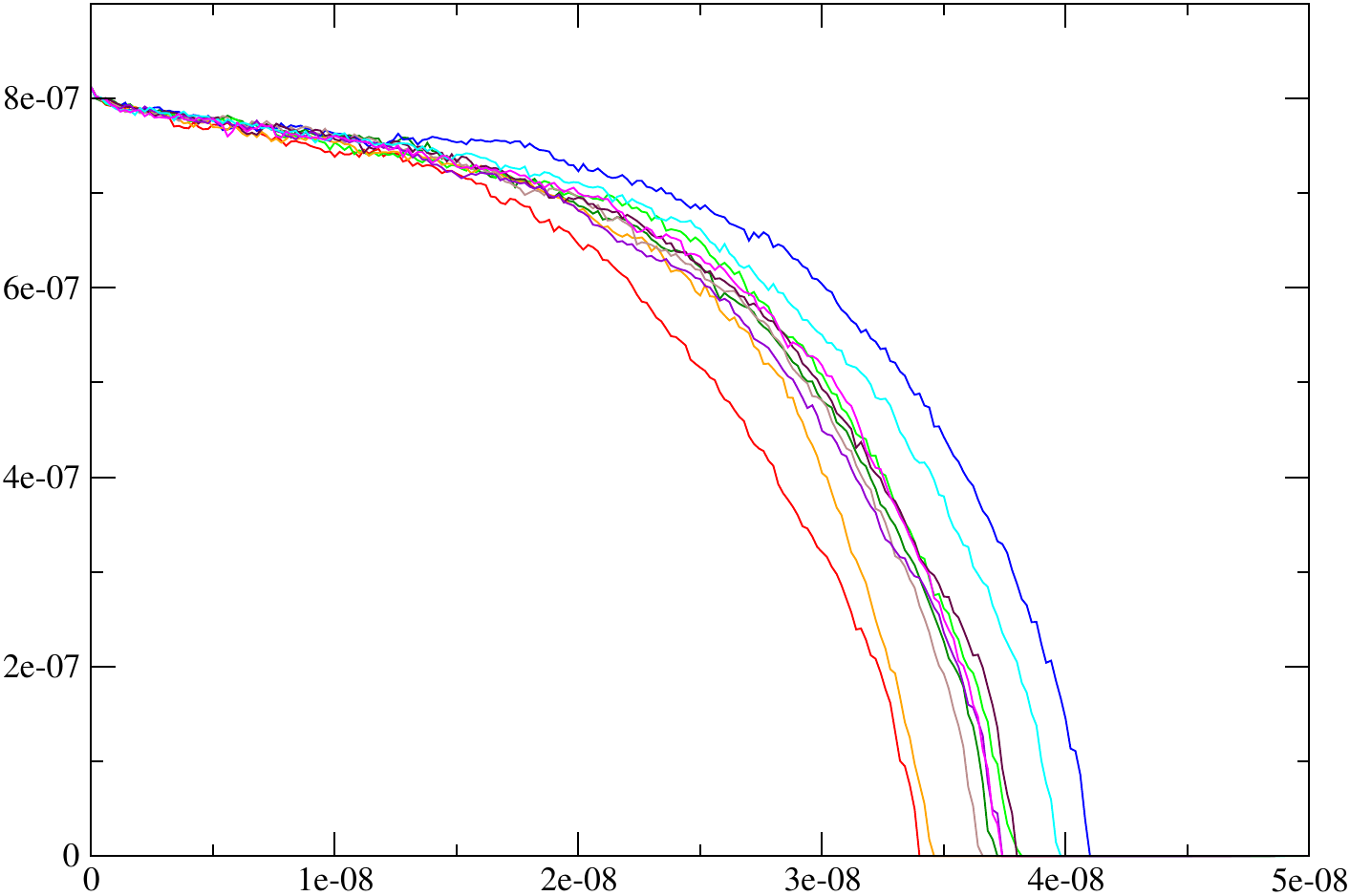}
    \includegraphics[width=0.3\textwidth]{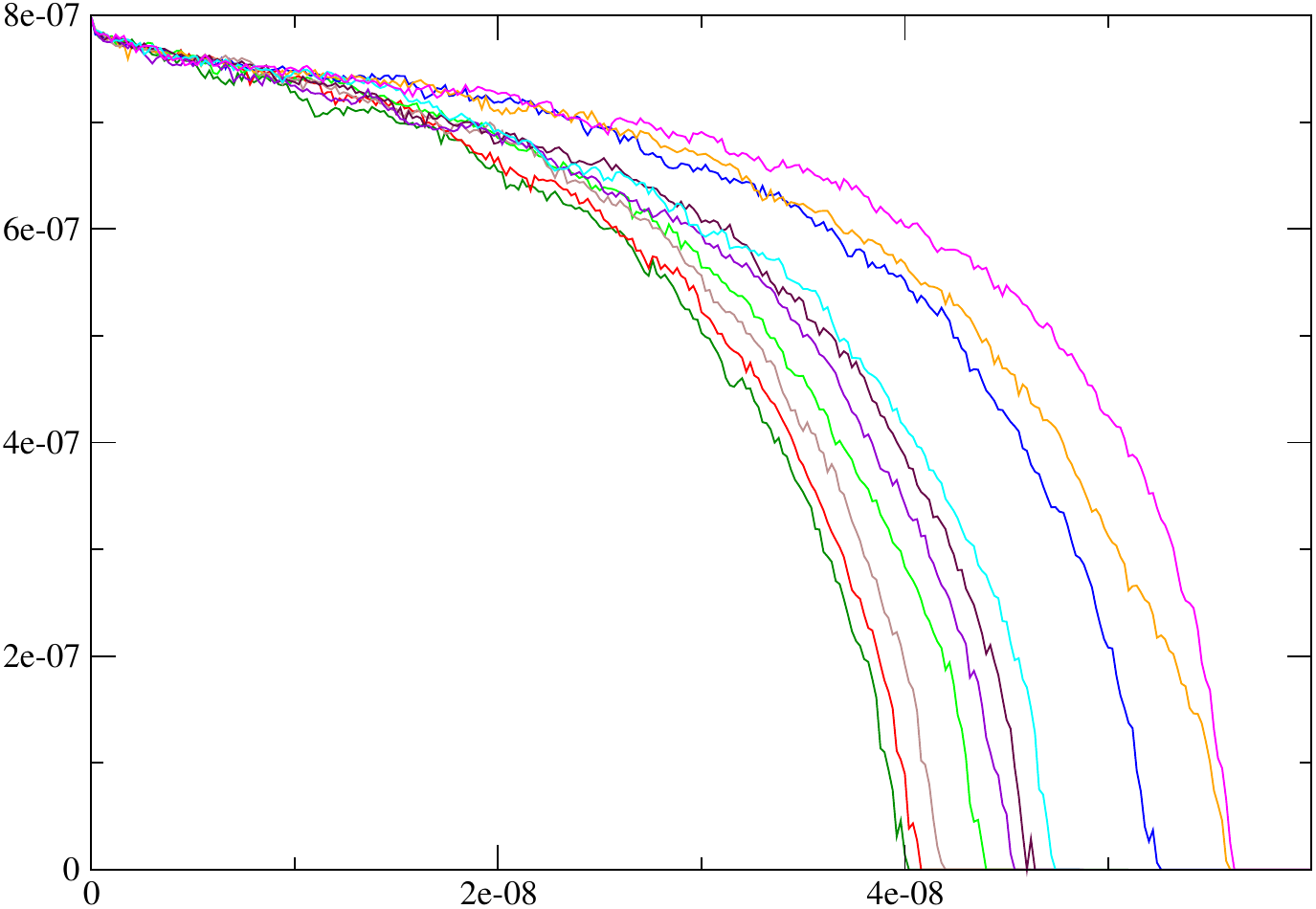}
    \includegraphics[width=0.3\textwidth]{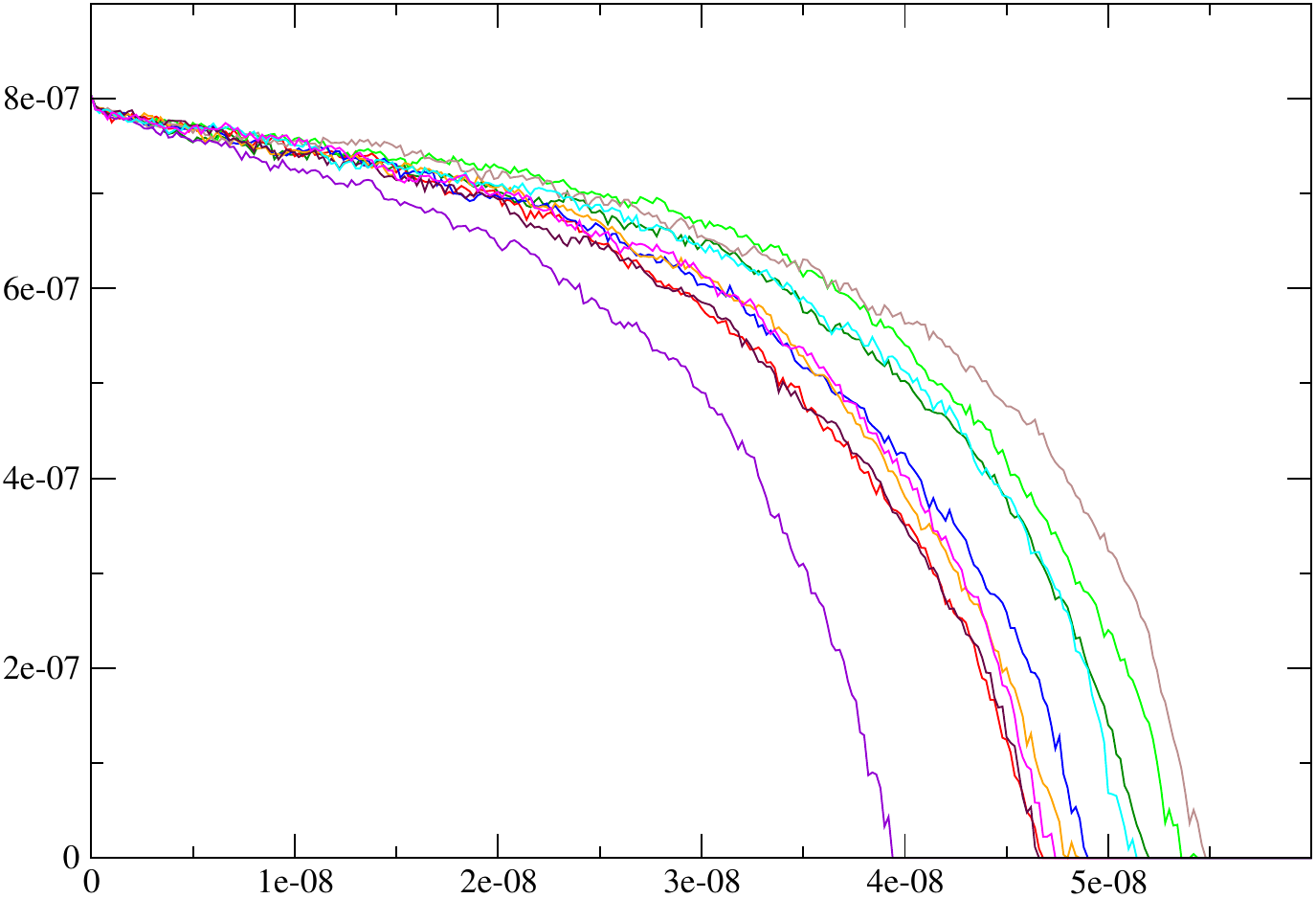}

    \caption{Minimum cylinder radius versus time in an ensemble of 10 stochastic simulations for (left) I-O, (middle) I-O*, (right) I-S-O. 
    }
    \label{fig:long_rmin}
\end{figure}

\begin{figure}
    \centering
    \includegraphics[width=0.3\textwidth]{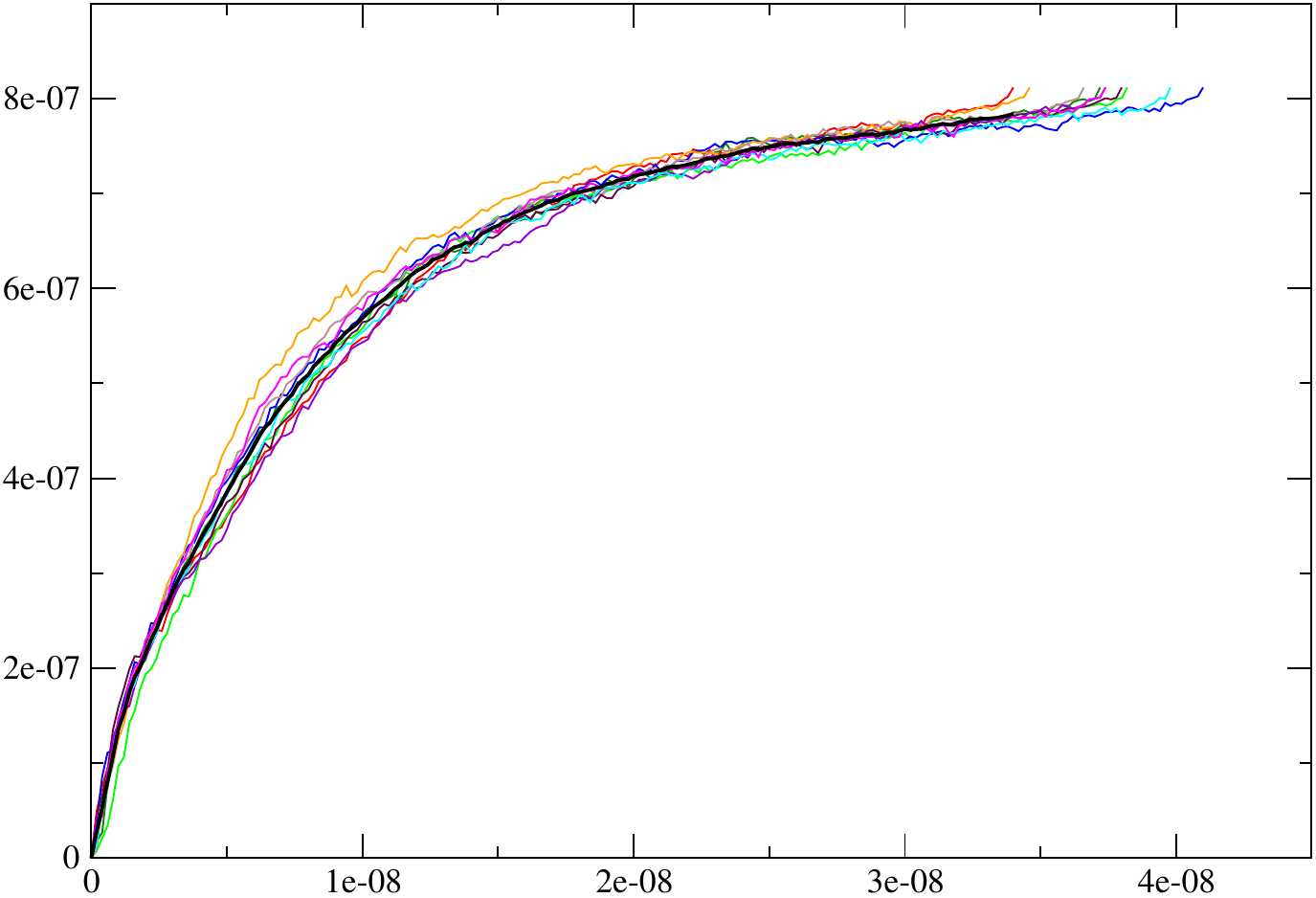}
    \includegraphics[width=0.3\textwidth]{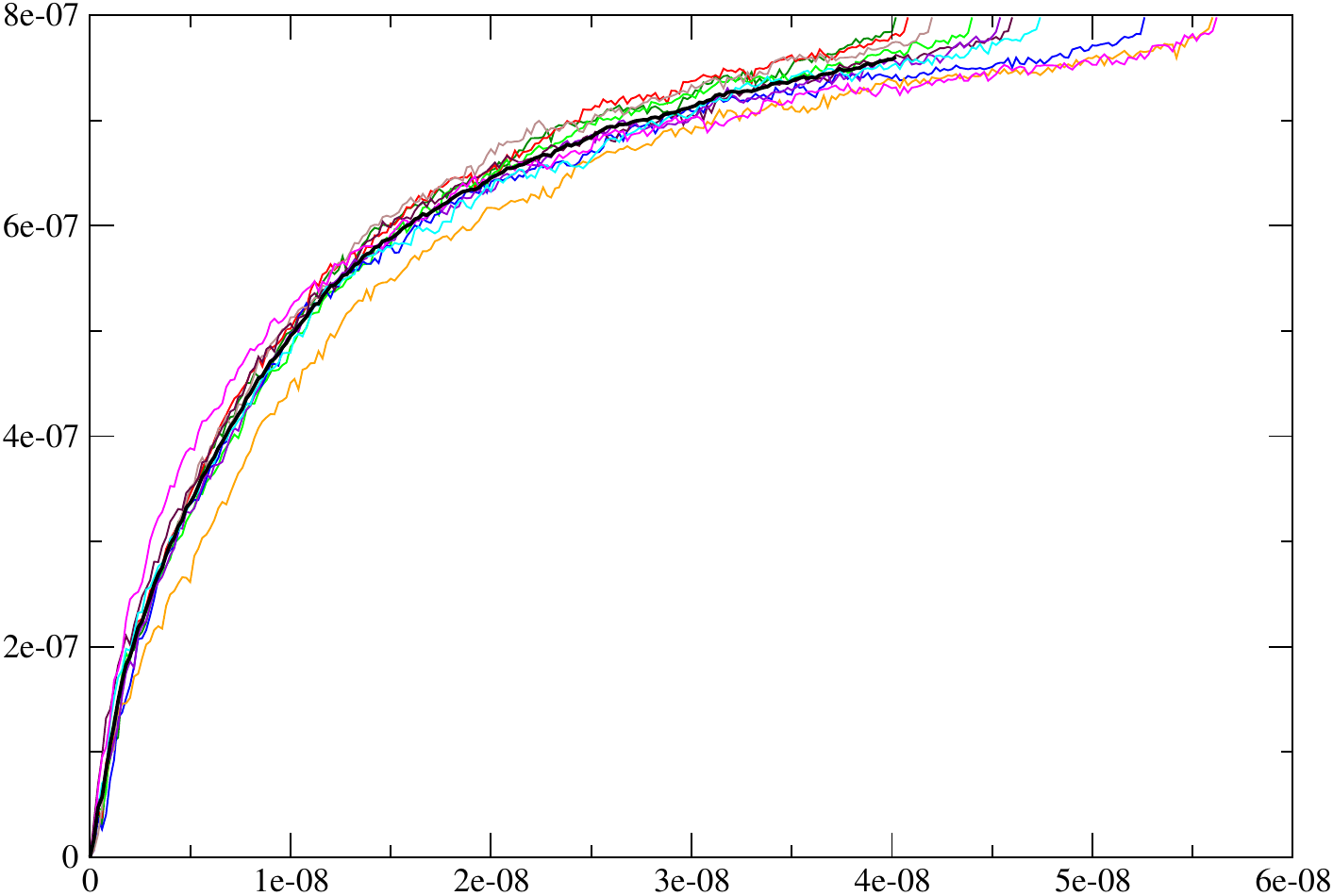}
    \includegraphics[width=0.3\textwidth]{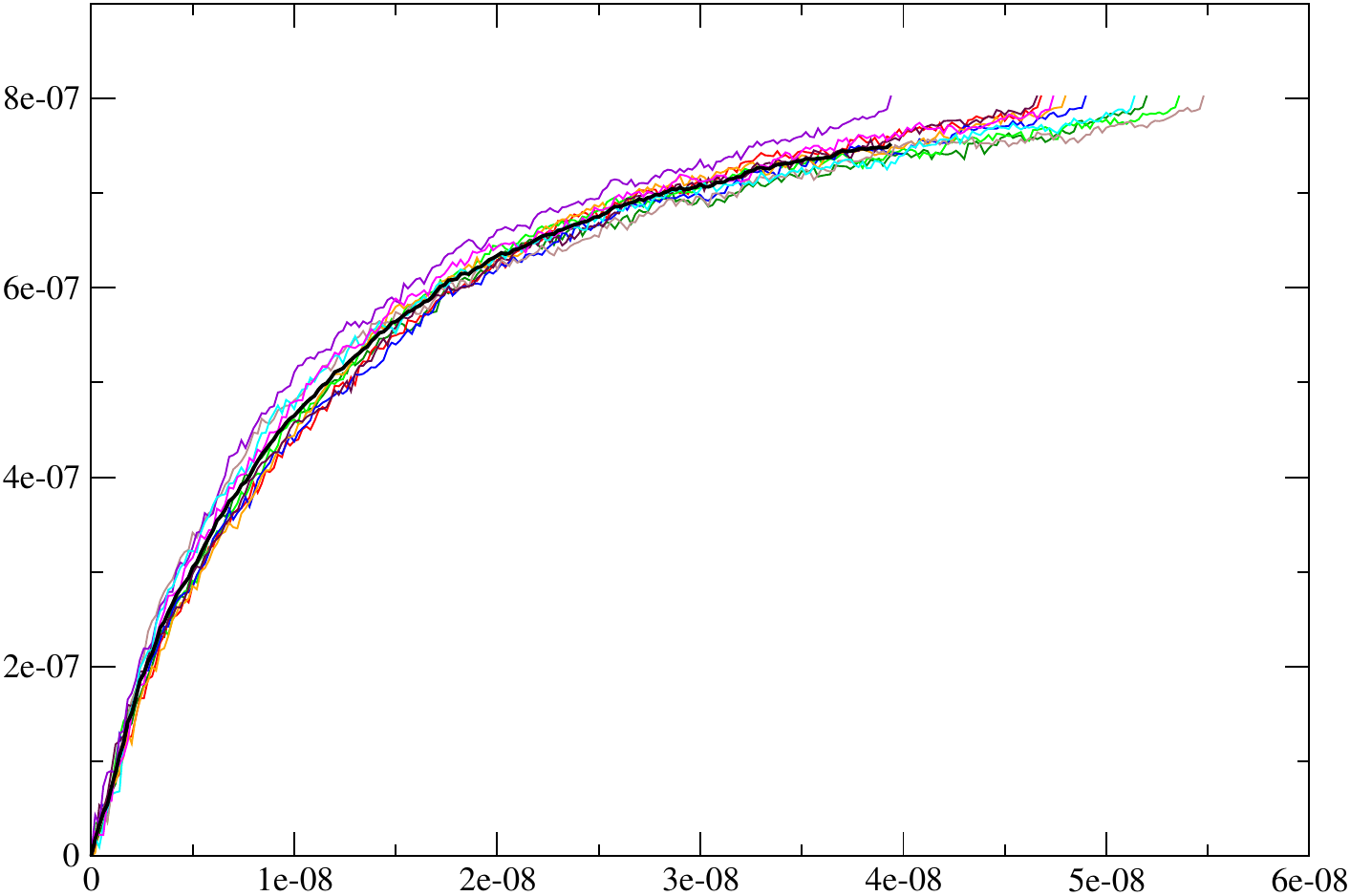}

    \caption{Minimum radius versus $\check{t} - t$ (time before pinch) for the data from Fig.~\ref{fig:long_rmin}.}
    \label{fig:long_rmin_rev}
\end{figure}

\begin{figure}
    \centering
        \includegraphics[width=0.5\textwidth]{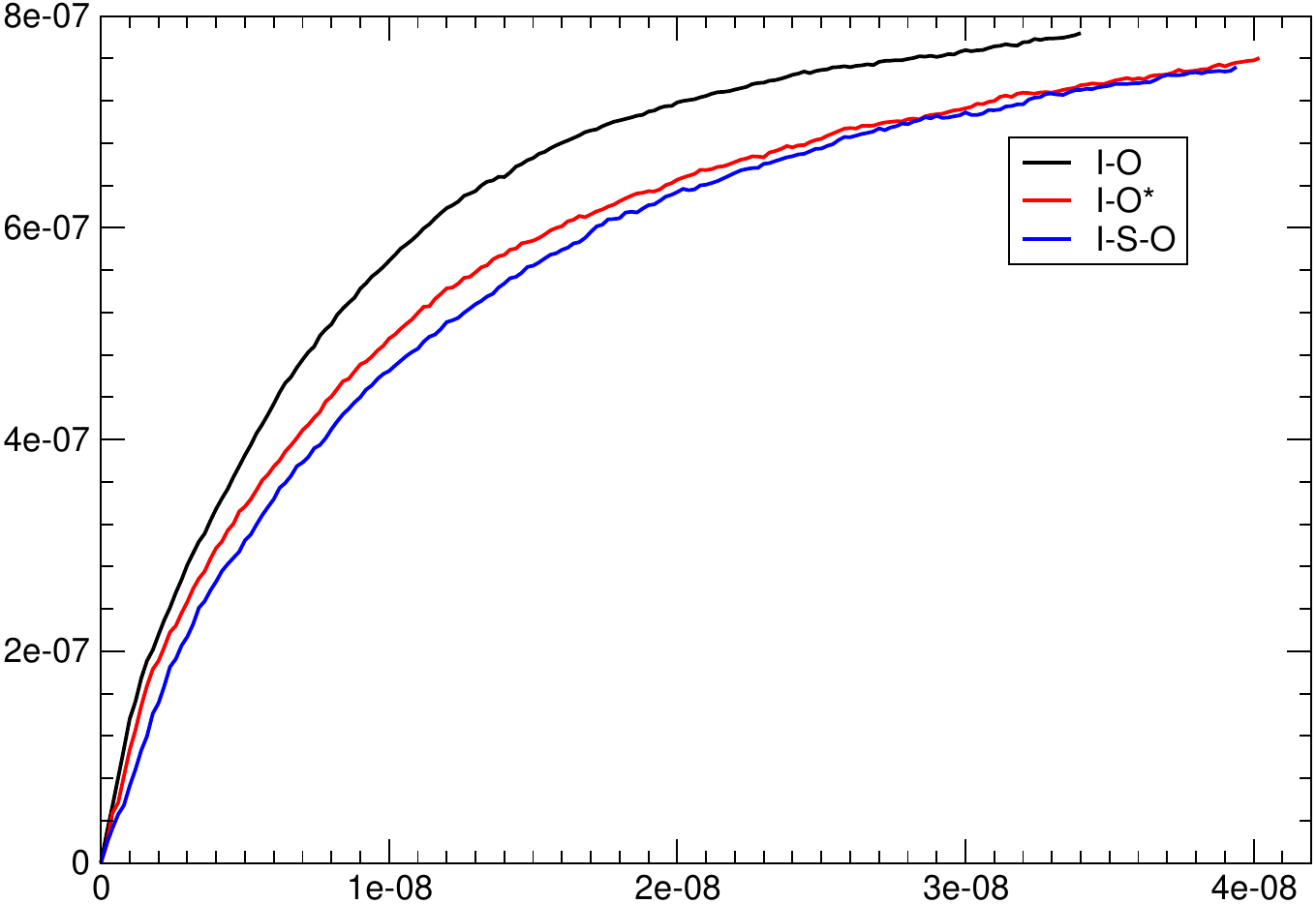}
        \caption{Ensemble average of $\check{t} - t$ (time before pinch) for the data from Fig.~\ref{fig:long_rmin_rev}.}
    \label{fig:long_rmin_rev_avg}
\end{figure}

Snapshots for stochastic simulations of long cylinders are shown in Figure~\ref{fig:stoch_long}.
In all three cases, the thermal fluctuations result in cylinders pinching into droplets sooner than was observed in the deterministic simulations.
Again, the I-O case pinches sooner than the I-O* case; however, when the I-O* and I-S-O cases are compared, it appears that the stabilizing effect of the surfactant is disrupted by thermal fluctuations.
These observations are confirmed in the results shown for $R_\mathrm{min}(t)$ in Figures~\ref{fig:long_rmin} and \ref{fig:long_rmin_rev}.
From these 10 run ensembles of runs, the average times for pinching for I-O, I-O*, and I-S-O cases are $\langle\check{t}\rangle = 37.42$ns, 47.06ns and 48.90ns with standard deviations of 2.10ns, 5.96ns, 4.42ns, respectively.
For all three mixture cases the average number of droplets formed was 3 indicating that the most unstable wavelength is $\lambda_\mathrm{max} \approx 11 \InitRad$. Small satellite droplets appeared in a few of the binary mixture (I-O and I-O*) runs but none appeared in the I-S-O runs. 
For the ensemble averaged data in Fig.~\ref{fig:long_rmin_rev_avg}, a fit of $R_\mathrm{min}$ between 1nm and 5nm gives: for I-O, $\check{a} = 0.6476$; for I-O*, $\check{a} = 0.6206$; and for I-S-O, $\check{a} = 0.6864$, which are similar to the exponents obtained in the deterministic simulations suggesting that the functional form of $R_\mathrm{min}(t)$ is not significantly affected by thermal fluctuations.

Finally, for each mixture case a single simulation was performed for a very long cylinder ($L = 1024$nm, $\mathcal{A}\approx 20$).  
The rupture time for these simulations was 31.6ns, 42.0ns, and 45.8ns for I-O, I-O* and I-S-O, respectively, which is somewhat lower than the mean for 256nm long cylinders. 
In the final configuration, I-O and I-S-O both had 13 droplets, whereas I-O* had 15; however, an ensemble of simulations would be needed to assess whether this is statistically significant.

\section{Concluding Remarks}\label{sec:Conclusions}

This paper presents a multispecies diffuse interface model within the fluctuating hydrodynamics framework, designed to simulate surfactant interfaces at the nanoscale. 
This model represents an extension of our previous work on binary mixtures to include ternary mixtures, where one species functions as a surfactant.
These miscible and immiscible species mixtures use a Cahn-Hilliard free energy density formulation combined with incompressible, isothermal fluctuating hydrodynamics in which the dissipative fluxes (like species diffusion and viscous stress) are the sum of deterministic and stochastic terms. 
The governing equations are formally written as stochastic partial differential equations (SPDEs), but they require discretization on a finite-sized mesh to introduce a high-wave number cutoff needed to obtain a well-defined mathematical model. 
Our formulation shares features with other deterministic multiphase models but offers a more generalized treatment of interfacial coefficients, making it suitable for modeling surfactant-laden interfaces. 

From numerical simulations of equilibrium systems, we demonstrated that for immiscible binary mixtures the measured surface tension obtained from Laplace pressure agrees well with Cahn-Hillard theory. 
For ternary mixtures containing a surfactant, a linear decrease in surface tension with increasing surfactant concentration was measured, consistent with Pockels' observations for surface-active compounds. 
This dependence on concentration results in Marangoni convection, where gradients in surface tension, often induced by non-uniform surfactant distributions, drive fluid flow. 
Analysis of thermal capillary wave spectra revealed that while the model generally agrees with classical capillary wave theory at low wavenumbers, the surfactant-laden mixture showed a significantly greater deviation from classical theory compared to two-species mixtures, presumably due to Gibbs elasticity. 
Although surfactants are known to damp capillary waves, this dynamic damping effect, caused by Marangoni forces, could not be directly observed from our measurements of the static structure factor; the analysis of the dynamic structure factor is a topic for future research.

In non-equilibrium simulations, we investigated the Rayleigh-Plateau instability and the spreading of surfactant patches, highlighting the impact of thermal fluctuations on their dynamics. 
Although deterministic simulations of short cylinders (circumference greater than length) were stable, thermal fluctuations in stochastic simulations induced pinching into droplets in all cases, with the surfactant having minimal effect. 
For long cylinders, deterministic simulations showed that higher surface tension accelerated the instability, and crucially, the Marangoni stress resulting from surfactant gradients significantly delayed the pinching process, in agreement with the general observation that surfactants have a stabilizing influence on thin films. 
However, in stochastic simulations of these long cylinders, thermal fluctuations led to earlier pinching, and the stabilizing effect of the surfactant observed in deterministic runs appeared to be disrupted. 
Similarly, during the spreading of surfactant patches driven by Marangoni convection, thermal fluctuations were found to partially suppress this convection, leading to smaller average peak heights in the propagating front.

The theoretical model and numerical methods presented here can be applied to a number of interesting phenomena. 
For example, the spread of a thin film laden with surfactants can result in a fingering instability.\cite{afsar2003fingering,warner2004fingering1,warner2004fingering2,craster2006fingering,chan2024fingering}
The interaction of droplets and bubbles, either by merger of neighboring particles or collisional impact, is strongly influenced by surfactants.\cite{rekvig2007molecular, baret2012surfactants, pan2016controlling, martin2015simulations, weheliye2017effect} 
Macroscopic studies of droplet impact and splashing on surfaces find that surfactants slow the dynamics of crown formation and collapse with Marangoni stresses leading to a taller and narrower crown compared to the surfactant-free case.\cite{quetzeri2025droplet}
It would be interesting to see how thermal fluctuations affect these surfactant-laden interface phenomena.

\begin{figure}
    \centering
   \includegraphics[width=0.85\textwidth]{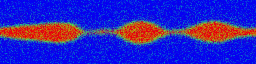}  \\ 
   \includegraphics[width=0.85\textwidth]{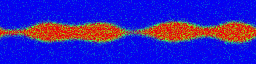} \\
 \includegraphics[width=0.85\textwidth]{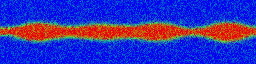}

    \caption{Snapshots at $t = 48$ns  from stochastic simulations of long ($L = 256$nm) cylinders for, top to bottom: I-S-O (baseline), I-S*-O (more miscible in I) and I-S**-O (less miscible in I) mixtures.}
    \label{fig:stoch_varS}
\end{figure}

Surfactant solubility plays an important role in the observed phenomena for surfactant-laden flows.
In the work reported here, we presented results for a partially soluble surfactant, yet by modifying the parameters in the free energy, $\boldsymbol{\chi}$ and $\boldsymbol{\kappa}$, the surfactant solubility can be increased or decreased.
In particular, the parameters can be adjusted so that the surfactant mixes more readily with the inner fluid (S*) or the surfactant is nearly immiscible with the inner fluid (S**).  
Figure \ref{fig:stoch_varS} shows snapshots of stochastic simulations with these differing surfactant properties.  
Preliminary results suggest that the immiscible surfactant increases the time to rupture compared to baseline surfactant case considered here so the interplay of surfactant solubility and thermal fluctuations merits future study.

\begin{acknowledgments}
Many of the numerical methods for fluctuating hydrodynamics used here were developed in the course of our long-term collaboration with our dear friend, Aleksandar Donev.
This work was supported by the U.S. Department of Energy, Office of Science, Office of Advanced Scientific Computing Research, Applied Mathematics Program under contract No. DE-AC02-05CH11231.
This research used resources of the National Energy Research Scientific Computing Center, which is supported by the Office of Science of the U.S. Department of Energy under Contract No. DE-AC02-05CH11231.
\end{acknowledgments}

\section*{Data Availability Statement}

The data that support the findings of this study are available from the corresponding author upon reasonable request.




\bibliography{Surfactant,TurbFHD}

\end{document}